**Distinct Modes of Functional Neural Organization in Autism: Insights from Dynamical Systems Analysis of Resting-State EEG**

Short title: Distinct Modes of Neural Organization in Autism

Sungwoo Ahn[1,2], Leonid L Rubchinsky[3,4], Evie A. Malaia[5*]

[1]Department of Mathematics, East Carolina University, Greenville, NC

[2]Center for Brain Stimulation, College of Allied Health Sciences, East Carolina University, Greenville, NC

[3]Department of Mathematical Sciences, Indiana University, Indianapolis, IN

[4]Stark Neurosciences Research Institute, Indiana University School of Medicine, Indianapolis, IN

[5]Department of Speech, Language and Hearing, University of Alabama, Tuscaloosa, AL

*Corresponding author: eamalaia@ua.edu

## Abstract

While differences in patterns of functional connectivity and neural synchronization have been reported between individuals on the autism spectrum and neurotypical peers at various age stages, these differences appear to be subtle and may not be captured by typical quantitative measures of EEG. We used the dynamical systems approach to analyze resting-state EEG to investigate fine-grained spatiotemporal organization of brain networks in autistic and neurotypical young adults. While power spectra showed minimal group differences, autistic participants exhibited higher Lyapunov exponents (indicating less stable neural dynamics), weaker phase synchronization, and lower clustering/efficiency of functional networks during eyes-open resting state, suggesting more random and less stably connected neural dynamics in comparison to those of neurotypical peers. Closing the eyes regularized neural dynamics in autistic but not neurotypical participants, with increases in synchrony strength, transient desynchronization patterning, and functional connectivity observed in the autistic group. The results point to the distinct modes of neural dynamics organization likely reflecting cumulative developmental adaptations to sensory inputs that shape both resting-state neural activity and cognitive processing strategies.

## 1. Introduction

Autism spectrum disorder (ASD) is a neurodevelopmental condition characterized by atypical patterns of behavior, communication, and social interaction. A growing body of research highlights the presence of altered functional connectivity and neural synchronization in individuals with ASD (Alotaibi & Maharatna, 2021; Trapani et al., 2022; Uddin et al., 2013). These differences in brain dynamics are thought to underlie the behavioral manifestations of the condition. Previous electroencephalography (EEG) studies have identified differences in functional connectivity and neural synchronization between individuals with ASD and typically developing peers (TD), including increased functional segregation and a higher desynchronization ratio in individuals with ASD (Malaia et al., 2016; 2020). These findings suggest that individuals with ASD exhibit altered patterns of neural communication and coordination, which may contribute to the behavioral and cognitive differences observed in their conditions. At the same time, the difference between TD and ASD brains does not markedly yield to standard quantitative measures of EEG (Li et al., 2022). Yet behaviorally, there are consistent features that are outcomes of information processing in the dynamical systems that differentiate autistic and neurotypical phenotypes (Davis & Crompton, 2021).

Consequently, there is a need to explore novel approaches that can provide a more comprehensive understanding of the neural underpinnings of ASD. Recent advances in complexity science and dynamical systems theory have provided access to new avenues for studying brain dynamics and neural synchronization. These approaches offer a powerful framework for examining the complex, multi-scale patterns of neural activity and their temporal structure, and could potentially reveal the intrinsic dynamics underlying unique neural functioning in ASD. The aim of this study is to use the dynamical systems approach in resting-state EEG data in autistic young adults. By adopting a dynamical systems perspective, we seek to provide insights into the functional connectivity patterns and neural synchronization dynamics that characterize ASD. Additionally, we aim to compare our findings with previous measures of synchrony and network properties in adolescents with ASD (Malaia et al., 2016, 2020) to contribute to the development of an integrative understanding of the neurodevelopmental trajectory of autism.

Resting state EEG data is a useful testing ground for investigating models of behavior of dynamical systems, especially in adults who have accumulated cognitive strategies to shape their own responses to the environment. In resting state EEG, both eyes-open and eyes-closed conditions merit investigation: while it is well-understood that closing the eyes potentiates alpha activity (visibly so over parieto-occipital regions), the oscillatory activity in the alpha range contributes non-trivially to several other measures such as network analysis. The topographical distribution of EEG parameters is also important to consider: while electrodes on the surface of the scalp do not allow source inferencing, they can help us understand the dynamical system behavior along the anterior-posterior axis (Barnea-Goraly et al., 2004; Beacher et al., 2012; Cheng et al., 2010; Ke et al., 2009) or left-right hemispheric balance (Herbert et al., 2005; Mak-Fan et al., 2012), both of which have been attested in autistic populations at the level of structure as well as function.

Prior EEG studies investigating neural activity in individuals with autism spectrum disorder (ASD) have produced mixed findings. While some report increased alpha or gamma power (e.g. Pierce et al., 2021), others have documented decreased or unchanged spectral power in these bands (Mathewson et al., 2012; Tierney et al., 2012). For instance, frontal alpha power has been found to both increase (Dickinson et al., 2018) and decrease (Coben et al., 2008) in ASD, depending on age, state , and methods used. A recent large-scale study by Barnes et al., (2025) found reduced functional connectivity in theta and alpha bands in children with ASD, with effects localized to the frontal cortex and not extending to global coherence. This finding aligned with ongoing concerns that traditional EEG metrics (which are usually based on averaged power or coherence) may fail to detect transient disruptions in neural connectivity. Thus, while further discrepancies among the results of the studies may arise from methodological variability (e.g., participant age, cognitive profiles, recording conditions), the fundamental limitation of conventional EEG analyses is that they rely heavily on time-averaged measures of spectral power and are not suitable for capturing the transient nature of dynamic neural activity.

To overcome these limitations, our study applies a dynamical systems framework to analyze resting-state EEG in autistic and neurotypical young adults. We focus on metrics like Lyapunov exponents, desynchronization patterning, and graph-based network efficiency, which are sensitive to both temporal and structural organization of brain activity. This approach extends recent methodological innovations (Barnes et al., 2025) by assessing not only functional connectivity strength but also its temporal persistence and topological coherence. Thus, our analysis goes beyond conventional measures to reveal latent dynamics that may underpin autistic cognition and behavior.

To comprehensively characterize the neural dynamics underlying ASD, we used a set of complementary measures. For those measures, where theoretical predictions existed (e.g., increased DR, reduced synchrony), we specified directional hypotheses; other measures were exploratory due to novelty or inconsistency in prior findings.

1) *band power measures:* prior studies have shown mixed results for specific band power measures in ASD, with reports of both increased and decreased power depending on frequency band, brain region, eyes closed or eyes open status, etc. (e.g., Tierney et al., 2012; Pierce et al., 2021). This metric is exploratory due to conflicting prior findings.

2) *fractal component (1/f) exponent:* this measure is expected to reflect the relative balance between random and oscillatory (functional) components in EEG, with flatter slopes (i.e. lower exponents) indicating increased oscillatory component contribution to EEG activity. Based on McCleod et al. (2024), we hypothesize that in individuals with ASD, the 1/f power-law exponent of resting-state EEG will be elevated in anterior regions and reduced in posterior regions.

3) *Lyapunov exponent:* this measure evaluates temporal stability of the EEG signal, with higher values indicating more chaotic dynamics of neural signal. Prior work suggests the autistic brain may show heightened dynamic instability (Barnes et al., 2025; Malaia et al., 2020); thus, we hypothesized that ASD participants' data will show higher Lyapunov exponents.

4) *Phase-locking index*: this measure quantifies the consistency of phase relationships between brain regions, representing functional connectivity independent of amplitude. ASD has been linked to reduced phase-based synchrony, especially in the frontal cortex and in theta/alpha bands (Barnes et al., 2025); thus, we expected to find reduced phase synchrony in young adults with ASD, especially in frontal regions.

5) *Desynchronization ratio (DR):* this measure quantifies the temporal structure of neural synchronization: higher DR suggests a system frequently shifting out of synchrony, which may underlie behavioral variability. Teenaged ASD participants have shown elevated DR (Malaia et al., 2020); we expected this pattern to be similar in young adults.

6) *Graph-theoretical measures (e.g., network efficiency; clustering):* these measures characterize the topology of functional brain networks, and global and local integration of signals. Previous work indicates some reduced efficiency and altered network integration in ASD youths (Malaia et al., 2016). We hypothesized that ASD participants show reduced network efficiency and clustering, especially in frontal subnetworks.

We further include the focus on both eyes open and eyes closed states. Although both are resting states, closing of the eyes provides a substantial perturbation to the state of brain networks, especially in autistic populations, which are characterized by sensory-defensive behaviors (Hazen et al., 2014). The approach of considering both the eyes-open (EO) and eyes-closed (EC) resting-state data effectively expands the dynamic range of the brain states at rest under consideration and may provide insight into autistic vs. neurotypical neural activity differences in situations without active inhibition of sensory inputs.

## 2. Methods

### 2.1. Participants

Two groups of young adults, undergraduate and graduate students at the University of Alabama, were recruited to participate in the study: a group of neurotypical (TD) young adults, and a group of autistic young adults (ASD). The participants, ages 18-30, were recruited from the students at the University of Alabama and UA-ACTS program for degree-seeking students with a medical diagnosis of ASD. All ASD participants were enrolled in the UA-ACTS program, which requires clinical ASD diagnosis supported by standardized tools such as the ADOS or ADI-R. Although we did not have direct access to individual diagnostic reports, we administered the Autism Quotient-10 (AQ-10, Allison et al., 2012) as an additional screening measure. All ASD participants met or exceeded the recommended threshold for clinical concern on the AQ-10, supporting the accuracy of diagnostic classification. Subcategories of ASD were not distinguished due to sample size limitations. The study received approval from the University of Alabama IRB. Participants provided written informed consent prior to beginning the study. Data was collected for 19 ASD individuals (10 F; age 18-27, $M = 20.7$, $SD = 1.9$), and 23 TD individuals (14 F; age 18-27, $M = 21$, $SD = 1.8$).

### 2.2. Procedure

#### *2.2.1. Behavioral measures*

To obtain behavioral measures of social anxiety and social anhedonia, the participants were asked to complete three questionnaires: the Autism Quotient (AQ, Baron-Cohen et al., 2001); the Perceived Report of Communication Apprehension (PRCA, McCroskey & McCroskey, 1988) was utilized to determine levels of social anxiety; and the Revised Social Anhedonia Scale was utilized to determine levels of social anhedonia (RSAS, Winterstein et al., 2011). Both scales have been used in autistic populations and have been demonstrated to be appropriate measures for the two specific dimensions of social functioning (Siew et al., 2017). The PRCA self-report questionnaire is composed of 24 statements concerning feelings about communication with others; the responders are asked to evaluate the degree to which each statement applies to them using a 5-point Likert scale. The PRCA has been shown to have strong content validity (McCroskey et al., 1985). The RSAS is another self-report scale appropriate for adults (Gadow & Garman, 2020); the 10-item version of the scale has strong internal consistency and mirrors effects found on the longer scale (Winterstein et al., 2011). The Autism Quotient (Baron-Cohen et al., 2001) questionnaire is normed for all populations – both clinical and neurotypical. Thus, while it was used as a confirmatory measure for autistic participants, it was also used to measure autistic traits in neurotypical participants. The shorter version of this scale, AQ-10, has been shown to have adequate validity as a measure of autistic traits (Lundin et al., 2019). Thus, it was utilized due to time constraints rather than the 50-item version.

*2.2.2. EEG recording procedure*

After participants provided informed, written consent as approved by the University of Alabama Institutional Review Board, they were fitted with high-density 64-channel EGI Hydrocel Geodesic Sensor net caps. Then, the participant was seated in a dark, quiet room. Participants were also instructed to minimize blinking and eye movements to reduce artifacts, and to relax while looking toward the turned-off task monitor, which served as a neutral visual anchor during the eyes-open condition. Continuous EEG data was collected during resting state conditions of eyes-open (EO, 5 minutes) and eyes-closed (EC, additional 5 minutes) using a Netstation EEG acquisition system. EEG data was sampled at 1000Hz/channel and referenced to the vertex during recording. Electrode impedances were kept below 50KOhm, consistent with the manufacturer's instructions.

## 2.3. Experimental data processing

### *2.3.1. Behavioral data*

On the autism screening assessment, the AQ-10, participants in the ASD group exhibited higher scores (AQ-10: $M = 6.8$, $SD = 2.1$) than the TD group (AQ-10: $M = 3$, $SD = 1.6$), [$t\ (40) = 6.4$, $p < .001$, *Cohen's d* =2]. On the social anxiety measure, the Perceived Report of Communication Apprehension (PRCA, McCroskey et al., 1985), participants in the ASD group exhibited higher scores ($M = 83.3$, $SD$=14) than the TD group ($M = 58.6$, $SD = 14$), [$t\ (40) = 5.6$, $p < .001$, *Cohen's d* = 1.7]. On the social anhedonia measure, the RSAS, participants in the ASD group exhibited higher scores ($M = 4.6$, $SD = 3.6$) than the TD group ($M = 0.26$, $SD = 0.54$), [$t\ (32) = 5.7$, $p < .001$, *Cohen's d* = 1.7]. These results confirm that there is a clear distinction in the behavioral characteristics of ASD and TD participants. ASD participants manifested increased social anxiety and social anhedonia, compared to TD participants. The ASD participants also scored higher on the AQ-10, which supports the previous diagnosis of autism.

*2.3.2. EEG data*

Data pre-processing was conducted on EGI Net Station software. In offline, the signal was filtered with a Butterworth Zero Phase Filter (high pass: 0.1 Hz, 48 dB/Oct; low pass: 60 Hz, 48 dB/Oct), and exported in edf format to Matlab software. The signals in clusters of interest were averaged to improve the signal-to-noise ratio, an accepted practice in the high-density recording literature (Brefczynski-Lewis et al., 2011). Areas of muscle activity, electrode drift, and eye-blink artifacts were mitigated through visual inspection and logistic infomax independent component analysis (RunICA) using the FieldTrip toolbox in MATLAB (Bell & Sejnowski, 1995; Delorme & Makeig, 2004; Oostenveld et al., 2011) for EEG/MEG analysis. Artifact components were identified based on their spatial topography and temporal characteristics of components, and subsequently removed from the data.

In this study, we further reduced the number of channels from 64-channel of EGI Hydrocel Geodesic Sensor to 18 EEG channels whose electrode positions are comparable to the standard 10-20 EEG system including FP1, FP2, F7, F3, FZ, F4, F8, T7, C3, C4, T8, P7, P3, P4, P8, O1, OZ, O2. The spatial distribution of 64 EEG channels is not uniform; therefore, we adapted the standard 10-20 EEG system, which provides a more evenly distributed layout and is widely recognized in both EEG research and clinical practice. Reducing the number of channels facilitates easier comparison and integration with many prior studies. The 18 standard EEG labels are evenly distributed across the scalp and offer nearly equal spacing between electrodes, ensuring comprehensive coverage of brain activity. EO and EC conditions were analyzed separately (Burnette et al., 2011). After deleting time intervals containing artifacts, some EEG signals had reduced time lengths from the original 5 min. Thus, the time durations in this analysis were 164.9 ± 84.9 sec for TD EO, 173.0 ± 97.1 sec for TD EC, 228.4 ± 98.3 sec for ASD EO, and 193.4 ± 103.7 sec (Mean ± SD) for ASD EC.

After removing artifacts in EEG signals, the recording durations varied across participants and conditions (EC, EO). To account for these differences and ensure systematics analysis, we segmented the artifact-free data into non-overlapping 30-sec windows. While some recordings contained only a few such windows, others contained several of these windows. This windowing approach allowed for consistent comparison across all conditions and groups. For further analysis, we used custom Matlab software to process the segmented EEG data. These 30 sec time windows give enough oscillatory cycles to compute the time-series measures even at low-frequency bands. All time-series measures below in Section 2.4 were computed for each individual electrode (18 channels) or each electrode pair (18*17/2=153 pairs) in 30 sec nonoverlapping time windows, and then each measure was averaged over the whole time period per electrode or per pair.

## 2.4. Data analysis

### *2.4.1. Spectral measures*

For the purposes of data analysis, the spectral bands were defined as follows: theta (4-7 Hz), alpha (8-12 Hz), beta (13-30 Hz), and low gamma (31-59 Hz). Power spectral density (PSD) from each electrode for each spectral band was computed using the built-in Matlab *pwelch* function for Welch's power spectral density estimate using 50% overlapped Hamming windows. Default parameters in Matlab were used including a window length of 256 samples (equivalent to 0.256 sec) and an overlap of 128 samples (0.128 sec) in the current study. This PSD is referred to here as the spectral power.

In addition to the spectral power analysis described above, we also utilized irregular-resampling auto-spectral analysis (IRASA; Wen & Liu, 2016) to separate fractal (1/f) and oscillatory components in the PSD of the signal. We further computed the power-law exponent of the 1/f component by fitting the log-log of the fractal components as in (Wen and Liu, 2016). The resulting characteristics are referred to here as the oscillatory power and the 1/f exponent. While the former is computed for each of the spectral frequency bands under consideration, the latter is, of course, not band-specific and is a broad-band characteristic of a signal as a whole.

### *2.4.2. Measure of dynamics regularity*

To characterize the regularity of neural dynamics we used the Lyapunov exponent, which characterizes the rate of exponential divergence of close trajectories in the reconstructed phase space. A larger Lyapunov exponent indicates a larger divergence rate, which accompanies the chaoticity of dynamics. We estimated the largest Lyapunov exponent using the built-in Matlab *lyapunovExponent* function (Abarbanel, 2012; Kantz & Schreiber, 2003) with time lag=10 ms and dim=5 and the expansion range=50 ms.

### *2.4.3. Synchronized dynamics measures*

The synchronization analysis methods used here were described previously in detail (Ahn et al., 2014; Ahn & Rubchinsky, 2013; Park et al., 2010). Briefly, signals were Kaiser windowed and digitally filtered using a finite impulse response filter in four frequency (theta, alpha, beta, low gamma) bands. The Hilbert transform was used to reconstruct phases of oscillations. The reconstructed phases were used to estimate the phase-locking index:

$$\gamma = \left\| \frac{1}{N} \sum_{j=1}^{N} e^{i\theta(t_j)} \right\|^2,$$

where $\theta(t_j)$ is the difference of the phases of oscillations at the given time point $t_j$. $N$ is the total number of such time points. This phase-locking index γ varies from 0 to 1 (perfect phase synchronization) and detects phase locking with any (not necessarily zero) phase lag (Hurtado et al., 2004; Pikovsky et al., 2001).

The phase-locking index γ characterizes the synchronization strength averaged over the analysis window. If the oscillations are synchronized on average, then it is possible to check whether oscillations are in the synchronized state at a specific cycle of oscillations (Ahn et al., 2011, 2014; Ahn & Rubchinsky, 2013,2020; Park et al., 2010). We extract intervals during which the phase difference is close to the preferred value and the intervals during which the phase difference substantially deviates from the preferred value (desynchronizations). Whenever the phase of one signal crosses zero level from negative to positive values, we record the phase of the other signal, generating a set of consecutive values $\{\phi_i\}$, $i=1,\ldots, M$. These $\phi_i$ represent the phase difference between two signals when the phase of one signal crosses zero and this is done once per cycle of oscillation. An example is illustrated in Fig. 1 AB, where we plot sines of the reconstructed phases for two simultaneously recorded signals. After determining the most frequent value of $\phi_i$, all the phases are shifted accordingly (for different episodes under consideration) so that averaging across different episodes (with potentially different phase shifts) is possible. Thus, this approach is not concerned with the value of the phase shift between signals, but rather with the maintenance of the constant phase shift (phase-locking).

Temporal dynamics are considered to be desynchronized if the phase difference deviates from the preferred phase difference by more than $\pi/2$ as in the earlier studies. The duration of the desynchronized episodes is measured in cycles of the oscillations, which allows for the comparison of temporal patterns of synchronization between different brain rhythms. Thus, in the data pictured in Fig. 1, the top row (Fig. 1A) illustrates the desynchronization episode lasting just one cycle of oscillations, while the middle row (Fig. 1B) illustrates the desynchronization episode lasting two cycles of oscillations (if the former events occur frequently, while the latter events occur rarely, the average phase-locking in the former case can be lower). This approach considers the maintenance of the phase difference in time and distinguishes between many short desynchronizations, few long desynchronizations, and possibilities in between even if they have the same average synchrony strength. This is quantified with the desynchronization ratio (DR): the ratio of the relative frequencies of the desynchronizations lasting for one cycle to longer than 4 cycles of oscillations (Ahn et al., 2014; Malaia et al., 2020). Smaller DR points to longer desynchronizations while larger DR points to shorter desynchronizations, see an example in the Fig. 1D,E. Temporal patterning of desynchronizations can vary independently of the average synchronization strength $\gamma$ (Ahn et al., 2014; Ahn & Rubchinsky, 2017; Nguyen & Rubchinsky, 2021, 2024) and may be more sensitive to the changes in the synchronous dynamics than changes in the synchronization strength (Ahn et al., 2018).

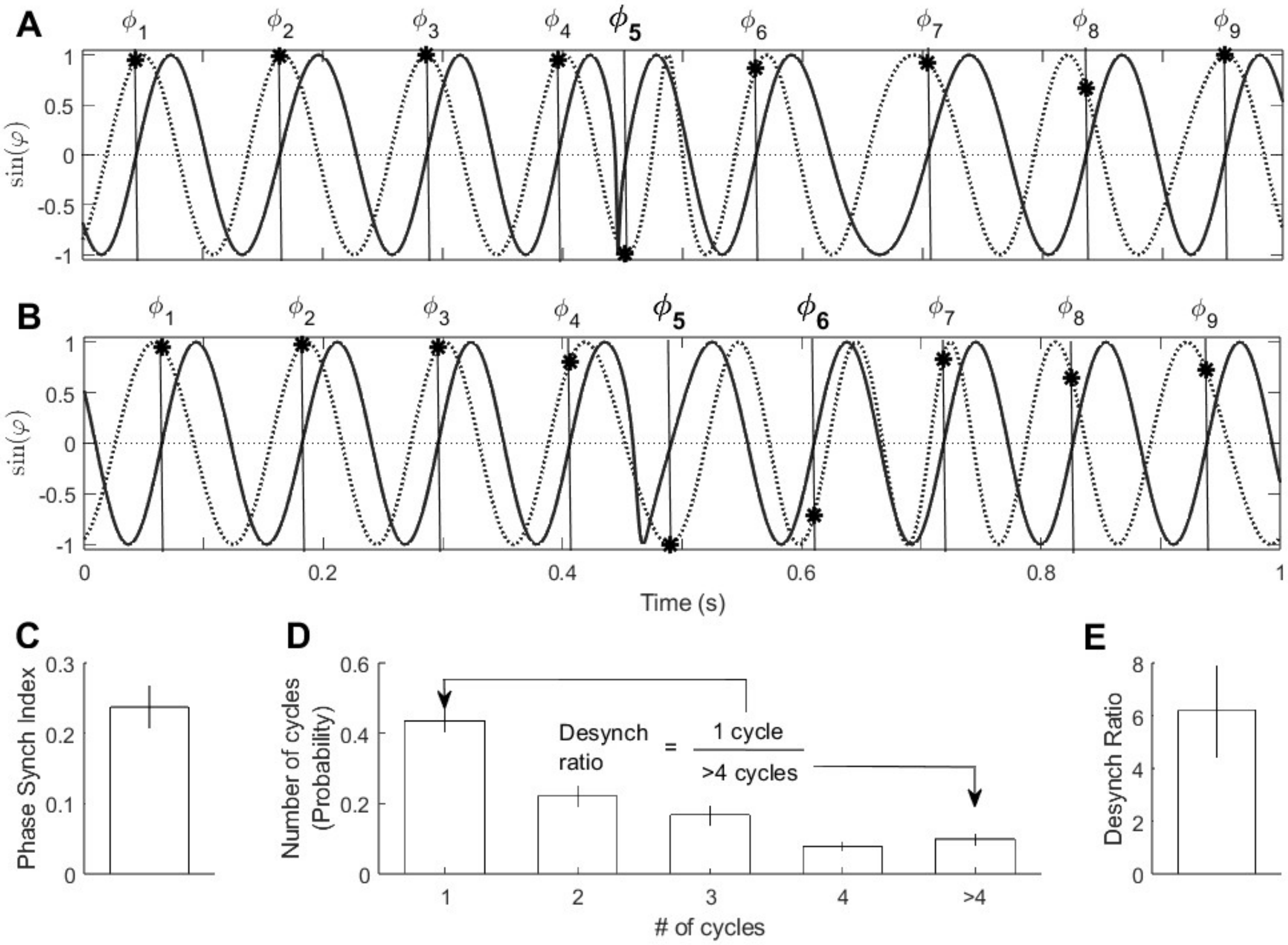

Fig 1. A, B are examples of the sines of the phases of recorded signals (solid and dotted lines) exhibiting desynchronization duration of one cycle of oscillations (A) and two cycles of oscillations (B) from a control subject with F3 and P3 channels during EO. The symbols represent the values of the phase of one signal, when the phase of the other signal crosses zero value on the way up (these events are marked by star symbols with vertical lines plotted at these values of time to help guiding the eye). C is the average synchronization index with Mean±SEM from the above subject. D is the distribution of desynchronization events durations with Mean±SEM. The desynchronization ratio is defined by the ratio of "1 cycle" bin over ">4 cycles" bin. E is the distribution of the desynchronization ratio corresponding to (D).

*2.4.4. Functional network measures*

We applied graph theory measures to networks defined as systems consisting of a set of nodes (representing electrodes) linked by edges (representing functional interactions). Initially, we constructed a symmetric connectivity matrix $A=[a_{ij}]$ (18 by 18 matrix) where the diagonal elements were set to zero. Here, each element $a_{ij}$ represents the phase-locking index γ between i-th and j-th nodes. To convert matrix A into a binary matrix B, we used a thresholding approach. In matrix B, a value of 0 represented no connection, and 1 represented a connection. We determined the threshold by using values of the phase-locking index γ for all pairs of electrodes (18*17/2 =153 pairs) from all control participants at each frequency band and at each condition. Specifically, we tested two different threshold values: the mean value or the mean value + 0.5 standard deviations of these phase-locking index from all control participants.

Two kinds of graph theory measures were calculated based on the binary matrix B. The clustering coefficient (CC) quantifies the tendency of nodes in a graph to form clusters or tightly interconnected groups. A higher clustering coefficient indicates a higher density of triangles or closed connections among connected nodes, reflecting a more segregated network structure (Bullmore & Sporns, 2012; Latora & Marchiori, 2001; Watts & Strogatz, 1998). The clustering coefficient of node *v* is calculated as the fraction of closed triangles among three nodes. Specifically, if *deg(v)* is the degree of a given node *v* and *T(v)* is the number of closed triangles centered at the node *v* then, *CC(v)* is defined as

$$CC(v) = \frac{T(v)}{\deg(v) \times (\deg(v) - 1)/2}$$

and it captures the density of connections and the tendency of the connectedness of *v* to be interconnected with other nodes. Global clustering coefficient *CC* is the average value of *CC(v)* across all the nodes of the network.

Furthermore, the characteristic of network efficiency is considered. The efficiency for each node assesses the connectedness of spatially distant regions. The efficiency for each node was computed as the mean value of the reciprocal shortest path lengths between the given node and all other nodes in the network. Specifically, if *d(v, u)* represents the shortest path length (minimum number of edges in any path) between node *v* and node *u,* then *EF(v)*, the efficiency of node *v*, is defined as the mean value of *1/d(v, u)* for $v \neq u \in V$, where *V* is the set of all nodes in the network. If there is no path from *v* to *u*, then *1/d(v, u)* is defined as zero. Global efficiency *EF* is the average value of *EF(v)* across all the nodes of the network. High efficiency may promote processing and effective information integration across the network.

These two measures may provide valuable insights into the brain's interconnectedness and the efficiency of information segregation and propagation through the network (Bullmore & Sporns, 2009, 2012; Latora & Marchiori, 2001; McDonnell et al., 2021; Micheloyannis et al., 2006; Stam, 2004; Watts & Strogatz, 1998).

*2.4.5. Statistical data analysis*

For each electrode (or each pair) per condition for each subject, we first computed the above neural dynamical measures (PSD, oscillatory power, 1/f exponent, Lyapunov exponent, γ, DR, CC, and EF) over every 30 sec non-overlapping windows (note that there were several of these 30 sec non-overlapping windows). Subsequently, each of these measures was averaged across all the non-overlapping windows.

To explore the effects of the locations, we created bilaterally-symmetric six regions of interest (ROIs) including channels in left anterior (LA; FP1, F3, F7), right anterior (RA; FP2, F4, F8), left central (LC; T7, C3), right central (RC; T8, C4), left parietal (LP; P7, P3, O1), and right parietal (RP; P8, P4, O2). By using six ROIs, one can reduce the dimensionality of the analysis as well as enhance statistical power. All comparisons were first subjected to a mixed design repeated measures of ANOVA (between-subject factor as a group; within-subject factors as a condition (EO and EC) and ROI) with significance set at $p<0.05$. Before using the mixed design or repeated ANOVA, sphericity was tested by using Mauchly's test. When the assumption of sphericity was violated, the degree of freedom in Greenhouse-Geisser correction was used. The one-way ANOVA is considered a relatively robust test against the normality assumption. Individual post-hoc two-sided t-tests (two-sample for group comparison or one-sample for condition comparison) were conducted with the significance level as 0.0083 in six ROIs analysis or 0.025 (whole brain consideration: two comparisons) (Bonferroni correction).

## 3. Results

### <u>3.1. Spectral measures</u>

#### *3.1.1. PSD*

There were no significant differences in the PSDs between ASD and TD groups in every region of interest (6 ROIs), in every frequency band, and in both conditions (eyes opened (EO) or eyes closed (EC)) (Supplemental Fig. 1). For both groups (ASD and TD), PSD changed as a function of closing the eyes. As the alpha band is potentiated (especially posteriorly) by eyes closing, the increase in alpha power was seen in both groups (ASD and TD), across all regions of interest (Fig. 2). There were additional changes in other frequency ranges. While there were not enough statistical interactions between Group, Condition, and ROI, we conducted exploratory investigations into how opening and closing eyes affected neural activity differently within each group and condition in each region. In TD, theta PSD during EC condition was increased in the right frontal region, as well as bilaterally posteriorly. In ASD, the theta PSD increase during EC condition was observed over the left posterior region only. Beta PSD increase during EC condition was observed in ASD over the left posterior region.

While there were not enough statistical interactions between Group and Condition for the whole brain consideration, we conducted exploratory investigations into how opening and closing eyes affected neural activity differently within each group and condition. There were no differences between groups in any of the frequency bands during eyes open, but closing the eyes led to some differences. In both TD and ASD, closing the eyes led to PSD increase at theta band, alpha band, and beta band (Fig. 2). Please refer to Table 1 (and Supplemental Table 1) for detailed statistical results.

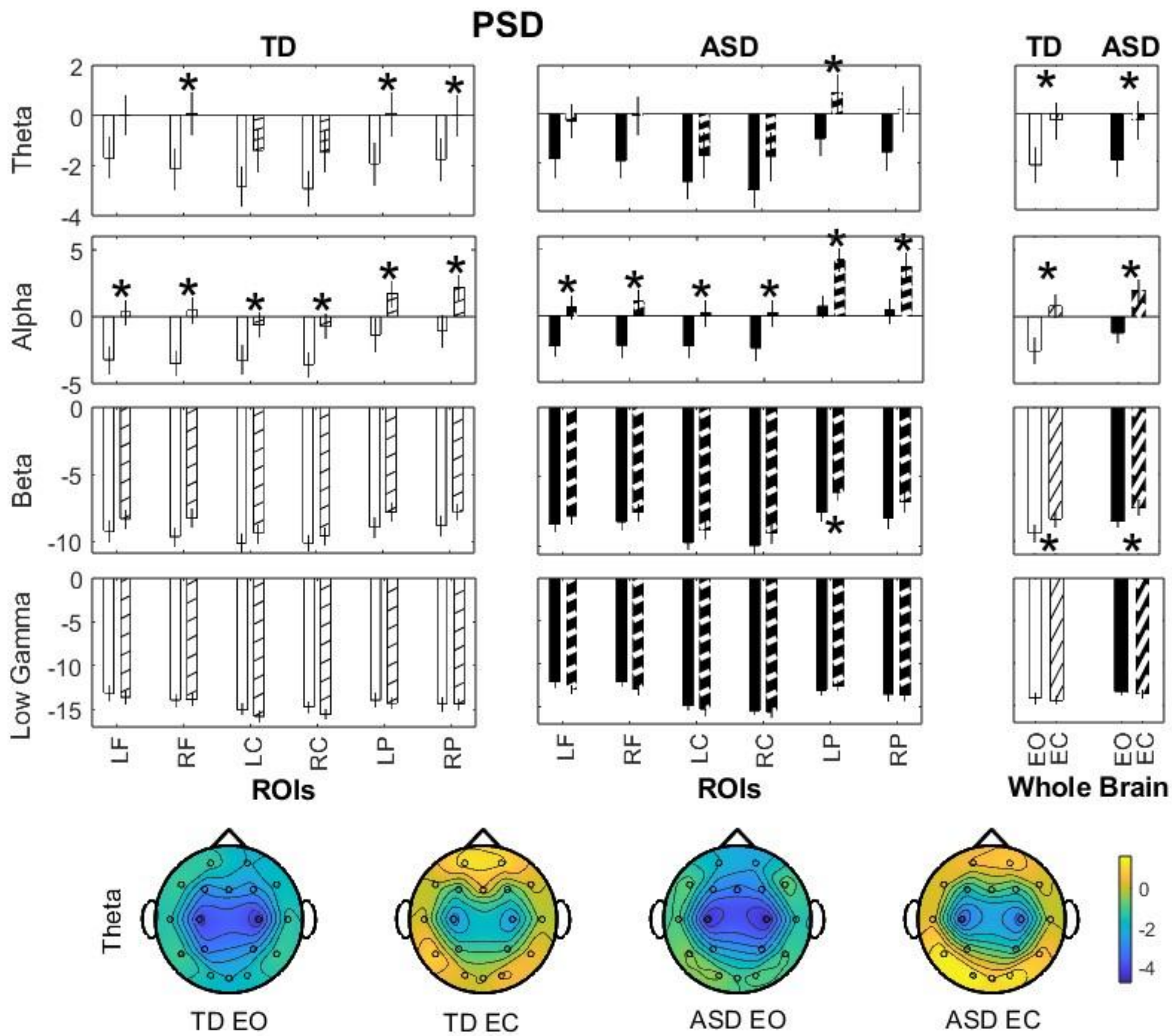

Fig 2. PSD. Condition comparison with TD (left panels) and ASD (right panels). TD EO (white bar), TD EC (white bar with black slanted lines), ASD EO (black bar), ASD EC (black bar with white slanted lines). Bar graphs with Mean ± SEM. *p<0.0083 in ROIs and *p<0.025 in Whole Brain. Topoplots for each group and each condition at theta band.

| | Condition | Group | ROI | Interaction |
|---|---|---|---|---|
| Theta Band | **F(1,40)=20.67, p=4.96e-5** | F(1,40)=0.01, p>0.05 | **F(3.1,124.6)=13.23, p=9.92e-8** | None (p>0.05) |
| Alpha Band | **F(1,40)=51.91, p=9.73e-9** | F(1,40)=0.99, p>0.05 | **F(2.8,113.6)=26.92, p=8.56e-13** | None (p>0.05) |
| Beta Band | **F(1,40)=12.00, p=1.28e-3** | F(1,40)=1.11, p>0.05 | **F(3.1,125.6)=12.09, p=3.36e-7** | None (p>0.05) |
| Low Gamma Band | F(1,40)=0.88, p>0.05 | F(1, 40)=1.00, p>0.05 | **F(3.1,122.2)=10.73, p=2.23e-6** | None (p>0.05) |

Table 1. ANOVA for PSD in ROI.

*3.1.2 Oscillatory power*

Oscillatory power presents outcomes similar to the PSD outcomes above. There were no significant differences between ASD and TD groups in every region, in every frequency band, and in both conditions (eyes opened or eyes closed) (Supplemental Fig. 2). While there were not enough statistical interactions between Group, Condition, and ROI, we conducted exploratory investigations into how opening and closing eyes affected neural activity differently within each group and condition in each region. There was a statistically significant broad alpha potentiation (all regions except left posterior) due to eyes closing, but only in ASD participants (Fig. 3).

While there were not enough statistical interactions between Group and Condition for the whole brain consideration, we conducted exploratory investigations into how opening and closing eyes affected neural activity differently within each group and condition. There were no differences between groups in any of the frequency bands and any condition (Supplemental Fig. 2). Closing the eyes led to higher oscillatory alpha power in both TD and ASD (Fig. 3). Please refer to Table 2 (and Supplemental Table 2) for detailed statistical results.

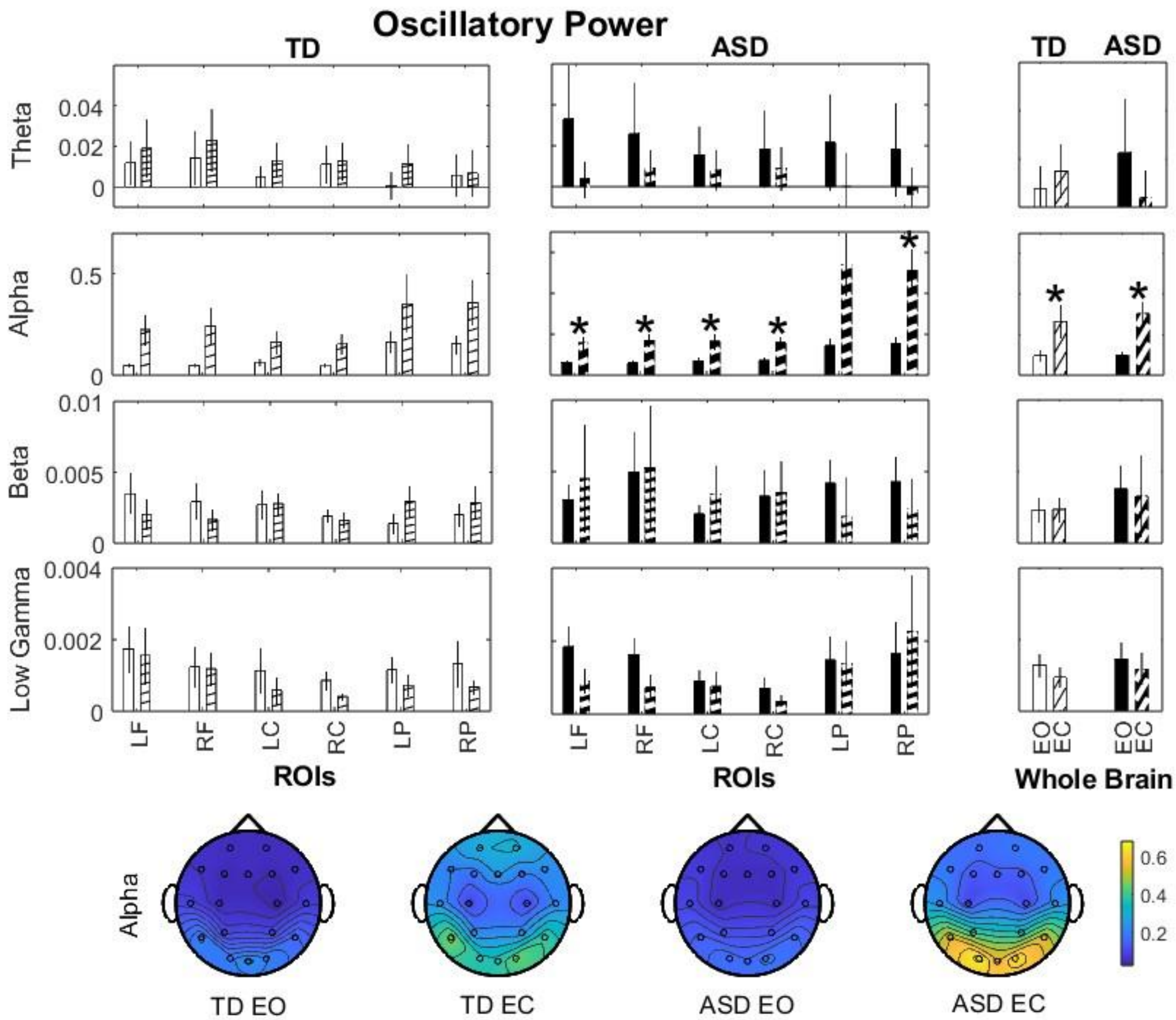

Fig 3. Oscillatory power. Condition comparison with TD (left panels) and ASD (right panels). TD EO (white bar), TD EC (white bar with black slanted lines), ASD EO (black bar), ASD EC (black bar with white slanted lines). Bar graphs with Mean ± SEM. *$p<0.0083$ in ROIs and *$p<0.025$ in Whole Brain. Topoplots for each group and each condition at alpha band.

| | Condition | Group | ROI | Interaction |
|---|---|---|---|---|
| Theta Band | F(1,40)=2.21, p>0.05 | F(1,40)=0.02, p>0.05 | F(1.7,76.9)=0.79, p>0.05 | None (p>0.05) |
| Alpha Band | **F(1,40)=17.00, p=1.83e-4** | F(1,40)=0.11, p>0.05 | **F(1.5,59.8)=14.36, p=4.92e-5** | **Cond*ROI [F(1.7,66.7)=5.71, p=7.94e-3]** |
| Beta Band | F(1,40)=2.05, p>0.05 | F(1,40)=3.58, p>0.05 | F(2.7,109.0)=0.65, p>0.05 | None (p>0.05) |
| Low Gamma Band | F(1,40)=3.49, p>0.05 | F(1,40)=1.93, p>0.05 | F(1.8,73.5)=1.24, p>0.05 | None (p>0.05) |

Table 2. ANOVA for Oscillatory Power in ROI.

*3.1.3. 1/f exponent*

There were no significant differences of 1/f exponents between ASD and TD groups in every region of interest and in both conditions (eyes opened or eyes closed) (Supplemental Fig. 3). While there were not enough statistical interactions between Group, Condition, and ROI, we conducted exploratory investigations into how opening and closing eyes affected neural activity differently within each group and condition in each region. Closing the eyes impacted the neural activity in each group differently. While there was no statistically significant change in the 1/f exponent in the TD group, the ASD group showed a significantly smaller 1/f exponent in all regions during opening the eyes compared to closing the eyes (Fig. 4).

While there were not enough statistical interactions between Group and Condition for the whole brain consideration, we conducted exploratory investigations into how opening and closing eyes affected neural activity differently within each group and condition. There were no significant differences between the ASD and TD groups in both conditions (Supplemental Fig. 3), but closing the eyes elevated the1/f exponent in ASD (but not TD) (Fig. 4). Please refer to Table 3 (and Supplemental Table 3) for detailed statistical results.

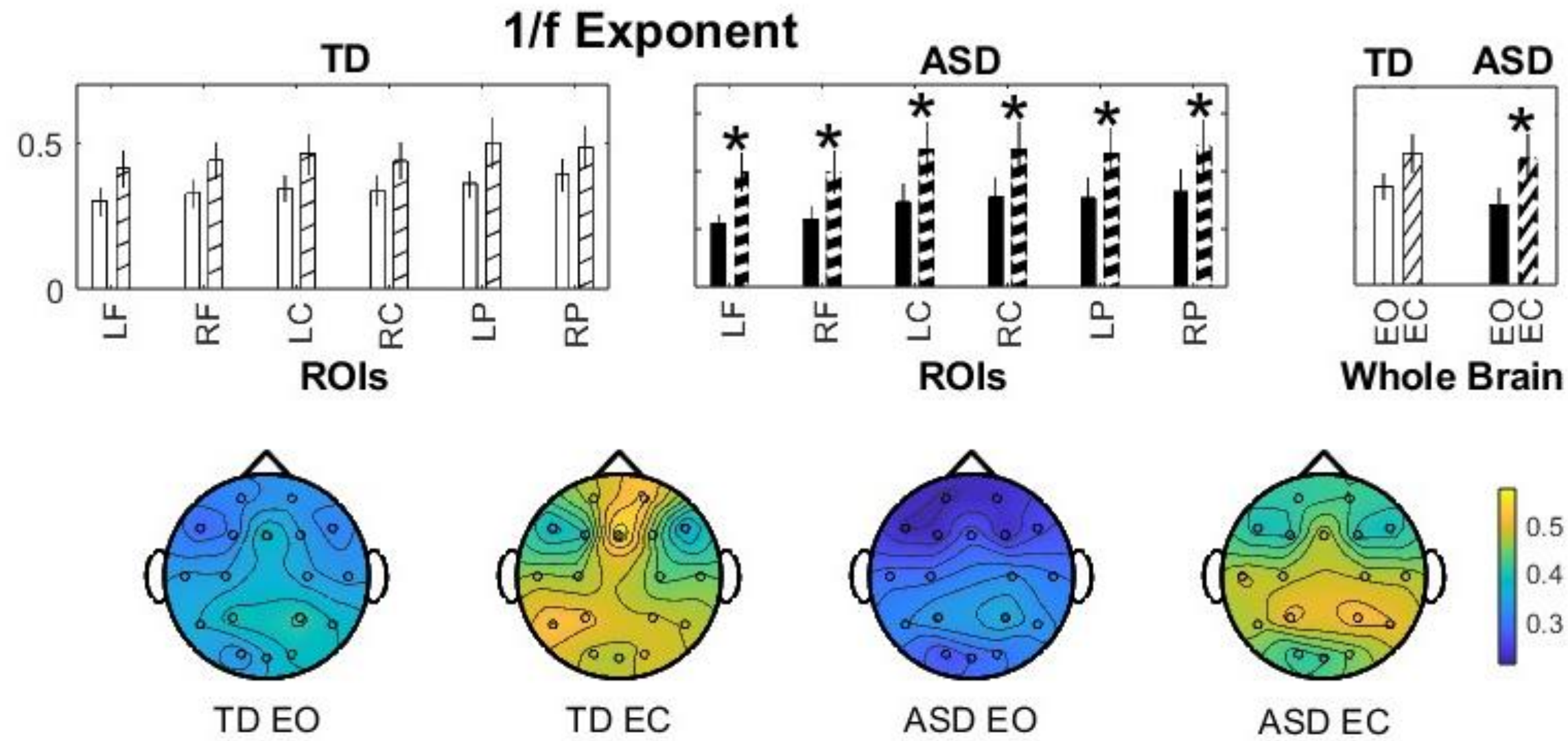

Fig 4. 1/f exponent. Condition comparison with TD (left panel) and ASD (right panel). TD EO (white bar), TD EC (white bar with black slanted lines), ASD EO (black bar), ASD EC (black bar with white slanted lines). Bar graphs with Mean ± SEM. *p<0.0083 in ROIs and *p<0.025 in Whole Brain. Topoplots for each group and each condition.

| Condition | Group | ROI | Interaction |
|---|---|---|---|
| **F(1, 40)=11.31, p=1.81e-3** | F(1, 40)=0.17, p>0.05 | **F(2.7,109.3)=8.18, p=1.01e-4** | None (p>0.05) |

Table 3. ANOVA for 1/f exponent in ROI.

## 3.2. Dynamics regularity measure

While there were not enough statistical interactions between Group, Condition, and ROI, we conducted exploratory investigations into how opening and closing eyes affected neural activity differently within each group and condition in each region. The Lyapunov exponent was higher in ASD than in TD in the EO condition, over the right central region of interest while there were no differences in any of the regions in the EC case (Supplemental Fig. 4). The effect of closing eyes elevates the Lyapunov exponent in TD (but not ASD) over left frontal, left posterior, and right central regions (Fig. 5).

While there were not enough statistical interactions between Group and Condition for the whole brain consideration, we conducted exploratory investigations into how opening and closing eyes affected neural activity differently within each group and condition. The effect of closing the eyes was to elevate the Lyapunov exponents in both TD group and ASD group (Fig. 5). Please refer to Table 4 (and Supplemental Table 4) for detailed statistical results.

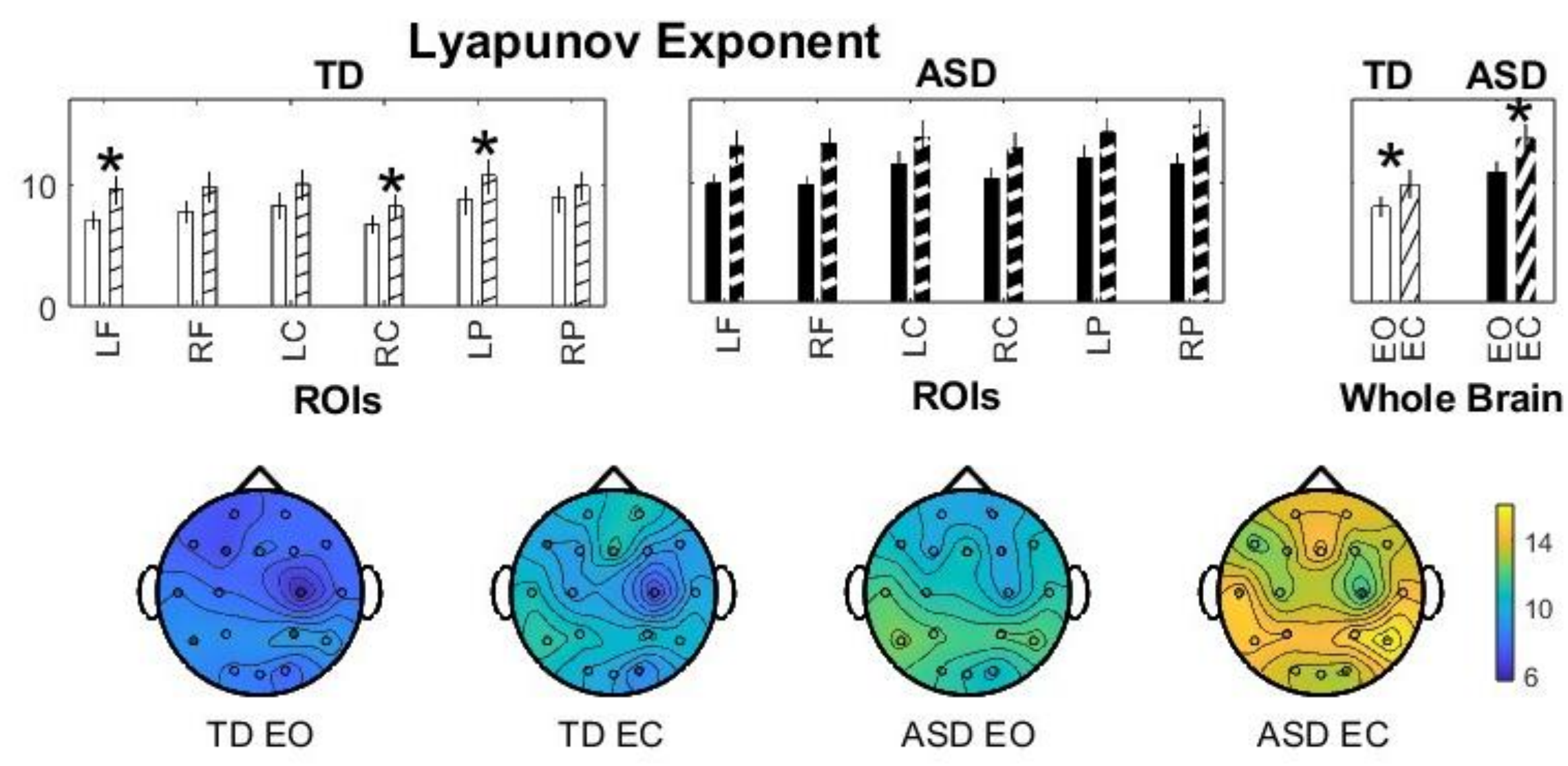

Fig 5. Lyapunov exponent. Condition comparison with TD (left panel) and ASD (right panel).

TD EO (white bar), TD EC (white bar with black slanted lines), ASD EO (black bar),

ASD EC (black bar with white slanted lines). Bar graphs with Mean ± SEM. *p<0.0083 in ROIs and *p<0.025 in Whole Brain. Topoplots for each group and each condition.

| Condition | Group | ROI | Interaction |
|---|---|---|---|
| **F(1, 40)=16.88, p=1.92e-4** | **F(1, 40)=5.91, p=1.97e-2** | **F(3.3, 132.7)=7.66, p=5.04e-5** | None (p>0.05) |

Table 4. ANOVA for Lyapunov exponent in ROI.

## 3.3. Synchronized dynamics measures

### *3.3.1. Phase synchronization index*

While there were not enough statistical interactions between Group, Condition, and ROI, we conducted exploratory investigations into how opening and closing eyes affected neural activity differently within each group and condition in each region. In the EO condition, the TD group demonstrated a significantly higher synchronization index (i.e. stronger phase synchronization) over the left central region at a low gamma band, as compared to the ASD group (no differences in the EC condition) (Supplemental Fig. 5). The effect of closing the eyes brought a statistically significant elevation of the phase locking index in ASD in alpha (broad anterior) and beta bands (left frontal) (Fig. 6). No statistically significant differences were observed for the TD group.

While there were not enough statistical interactions between Group and Condition for the whole brain consideration, we conducted exploratory investigations into how opening and closing eyes affected neural activity differently within each group and condition. During the EO, TD showed higher phase synch than ASD at the low gamma band (Supplemental Fig. 5). Closing the eyes led to significant elevation of the phase locking index in ASD (but not in TD) at alpha band and beta band (Fig. 6). Overall, the values of the phase locking index were in the 0.1-0.3 range. Please refer to Table 5 (and Supplemental Table 5) for detailed statistical results.

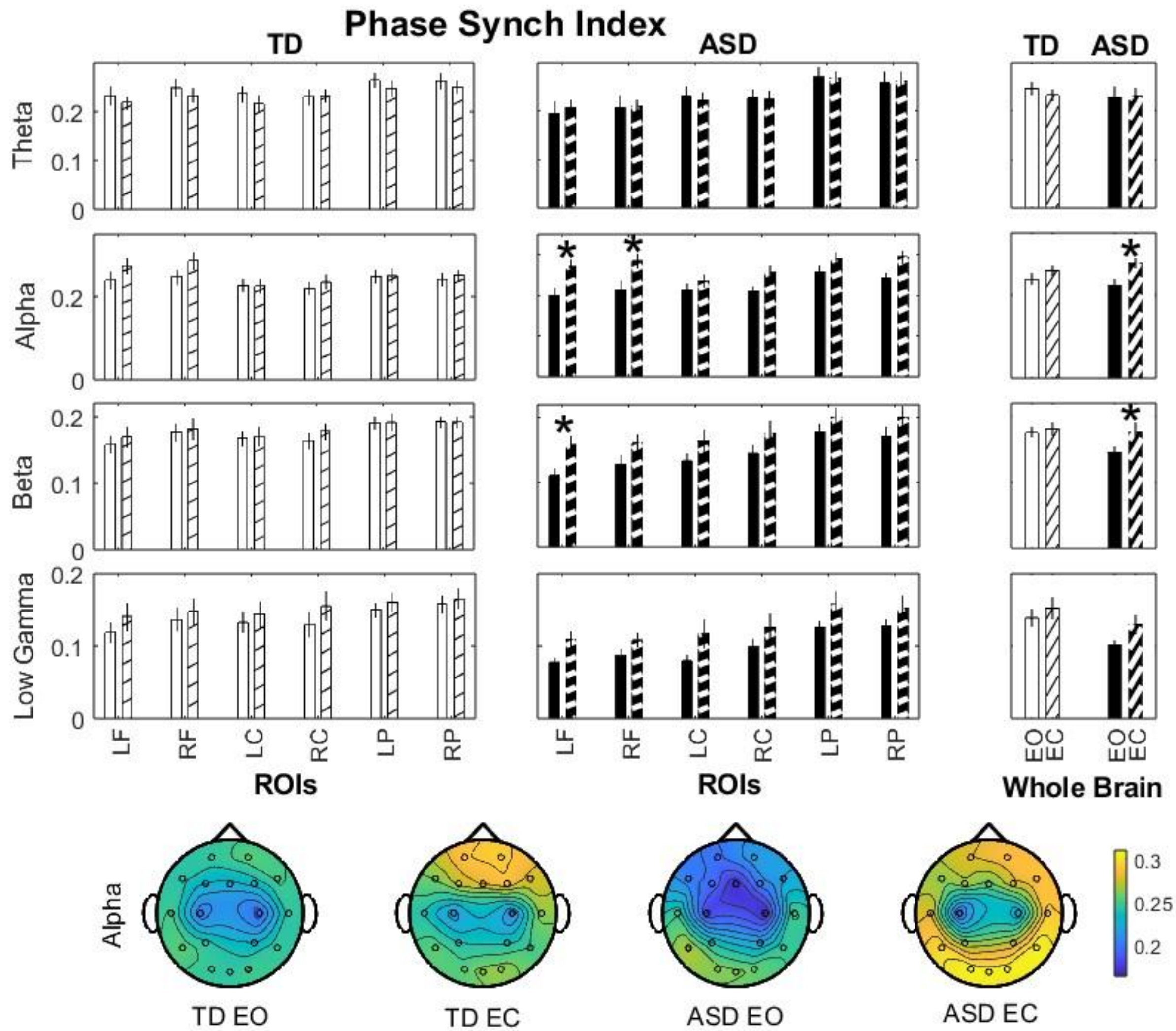

Fig 6. Phase synchronization index. Condition comparison with TD (left panels) and ASD (right panels). TD EO (white bar), TD EC (white bar with black slanted lines), ASD EO (black bar), ASD EC (black bar with white slanted lines). Bar graphs with Mean ± SEM. *p<0.0083 in ROIs and *p<0.025 in Whole Brain. Topoplots for each group and each condition at alpha band.

| | Condition | Group | ROI | Interaction |
|---|---|---|---|---|
| Theta Band | F(1,40)=0.31, p>0.05 | F(1,40)=0.14, p>0.05 | **F(3.7,147.2)=12.65, p=2.05e-8** | None (p>0.05) |
| Alpha Band | **F(1,40)=13.24, p=7.77e-4** | F(1,40)=0.01, p>0.05 | **F(3.5,138.6)=5.27, p=1.04e-3** | **Cond*ROI [F(3.5,139.0)=3.58, p=1.15e-2]** |
| Beta Band | **F(1,40)=6.20, p=1.70e-2** | F(1,40)=1.51, p>0.05 | **F(2.8,112.8)=9.13, p=2.83e-5** | None (p>0.05) |
| Low Gamma Band | **F(1,40)=5.89, p=1.98e-2** | F(1,40)=3.91, p>0.05 | **F(2.9,114.1)=8.69, p=4.23e-5** | None (p>0.05) |

Table 5. ANOVA for Phase synchronization index in ROI.

*3.3.2. Desynchronization ratio*

While there were not enough statistical interactions between Group, Condition, and ROI, we conducted exploratory investigations into how opening and closing eyes affected neural activity differently within each group and condition in each region. The desynchronization ratio was significantly higher in TD, when compared with the ASD group, in the low gamma band over the left central regions in EO condition (Supplemental Fig. 6). Closing the eyes leads to significant elevation of DR in ASD (but not TD) at the alpha band in anterior regions bilaterally (Fig. 7).

While there were not enough statistical interactions between Group and Condition for the whole brain consideration, we conducted exploratory investigations into how opening and closing eyes affected neural activity differently within each group and condition. During the EO, there were significant differences between ASD and TD at the low gamma band (Supplemental Fig. 6). Closing the eyes leads to a significant elevation of DR in ASD (but not in TD) at the alpha band (Fig. 7). Please refer to Table 6 (and Supplemental Table 6) for detailed statistical results.

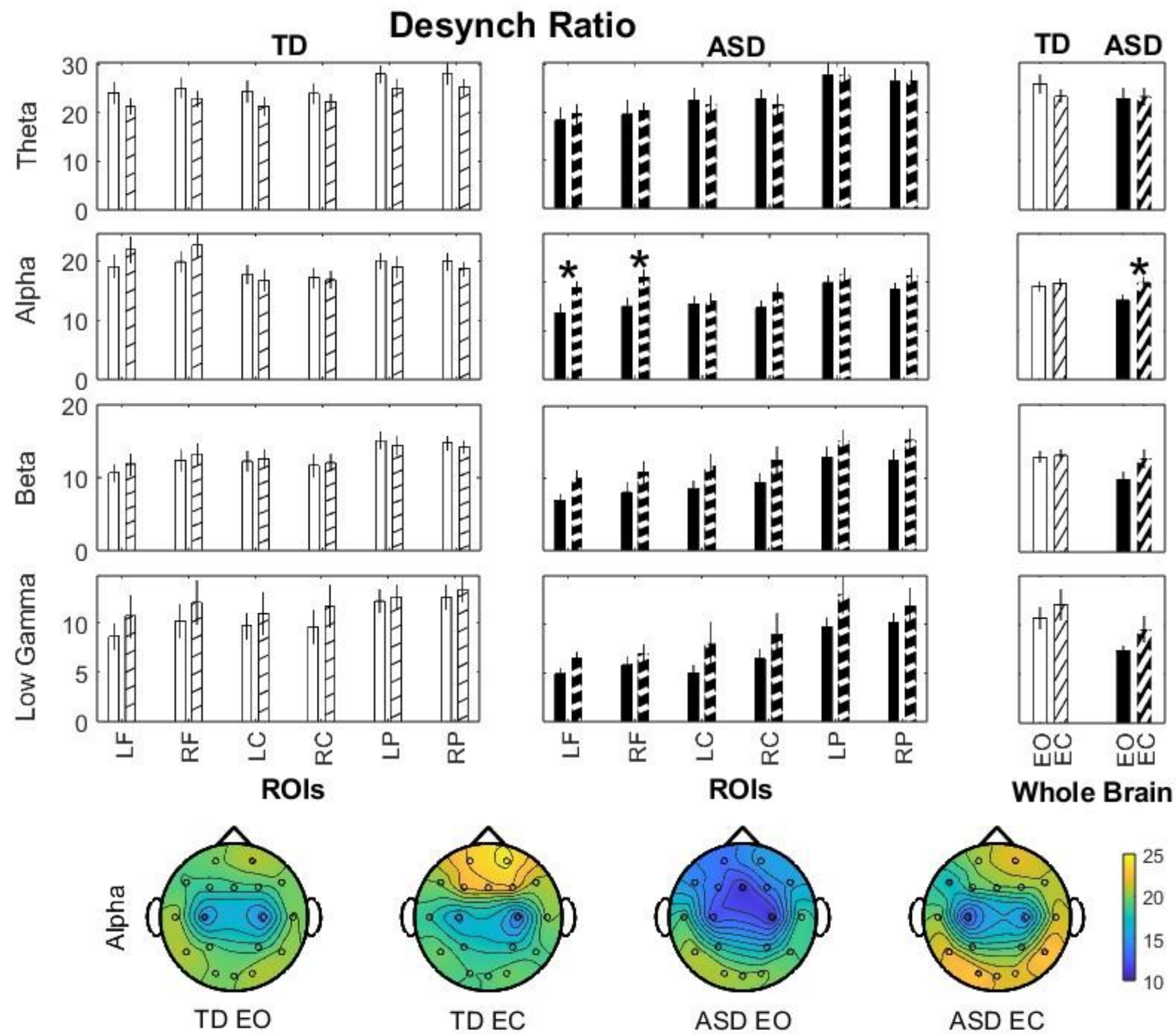

Fig 7. Desynchronization ratio (DR). Condition comparison with TD (left panels) and ASD (right panels). TD EO (white bar), TD EC (white bar with black slanted lines), ASD EO (black bar), ASD EC (black bar with white slanted lines). Bar graphs with Mean ± SEM. *p<0.0083 in ROIs and *p<0.025 in Whole Brain. Topoplots for each group and each condition at alpha band.

| | Condition | Group | ROI | Interaction |
|---|---|---|---|---|
| Theta Band | F(1,40)=0.82, p>0.05 | F(1,40)=0.40, p>0.05 | **F(3.6,141.9)=13.35, p=1.39e-8** | None (p>0.05) |
| Alpha Band | F(1,40)=3.48, p>0.05 | F(1,40)=0.70, p>0.05 | **F(3.5,139.7)=4.38, p=3.57e-3** | **Cond*ROI [F(3.4,136.3)=3.15, p=2.21e-2]** |
| Beta Band | **F(1,40)=5.00, p=3.10e-2** | F(1,40)=1.78, p>0.05 | **F(2.8,111.7)=8.67, p=5.08e-5** | None (p>0.05) |
| Low Gamma Band | F(1,40)=3.76, p>0.05 | F(1,40)=3.61, p>0.05 | **F(2.7,106.3)=8.44, p=9.19e-5** | None (p>0.05) |

Table 6. ANOVA for Desynchronization ratio (DR) in ROI.

## 3.4. Functional network measures

### *3.4.1. Clustering coefficient*

Two values of the threshold were considered (Section 2.4.4). For the threshold value equal to the mean of the phase synchronization index $\gamma$ of all pairs of electrodes from all control participants, there were no differences between TD and ASD in any of the regions in any of the frequency bands in any condition (Supplemental Fig. 7). Closing the eyes yielded no statistically significant differences either within the group (Supplemental Fig. 8).

While there were not enough statistical interactions between Group and Condition for the whole brain consideration, we conducted exploratory investigations into how opening and closing eyes affected neural activity differently within each group. During EO condition, CC was higher in TD than ASD at a low gamma band (Supplemental Fig. 7).

We considered a higher, (and thus more discriminating) threshold value equal to the mean + 0.5SD of $\gamma$. While there were not enough statistical interactions between Group, Condition, and ROI, we conducted exploratory investigations into how opening and closing eyes affected neural activity differently within each group and condition in each region. CC was higher in TD than ASD in EO condition over the right frontal region at both the theta and beta bands (Fig. 8). Closing the eyes elevated CC in ASD (but not in TD) at the alpha band over the left frontal region (Supplemental Fig 9).

While there were not enough statistical interactions between Group and Condition for the whole brain consideration, we conducted exploratory investigations into how opening and closing eyes affected neural activity differently within each group. Closing the eyes also elevated CC in ASD (but not in TD) at alpha band (Supplemental Fig 9). Please refer to Table 7 (and Supplemental Table 7,8,9) for detailed statistical results.

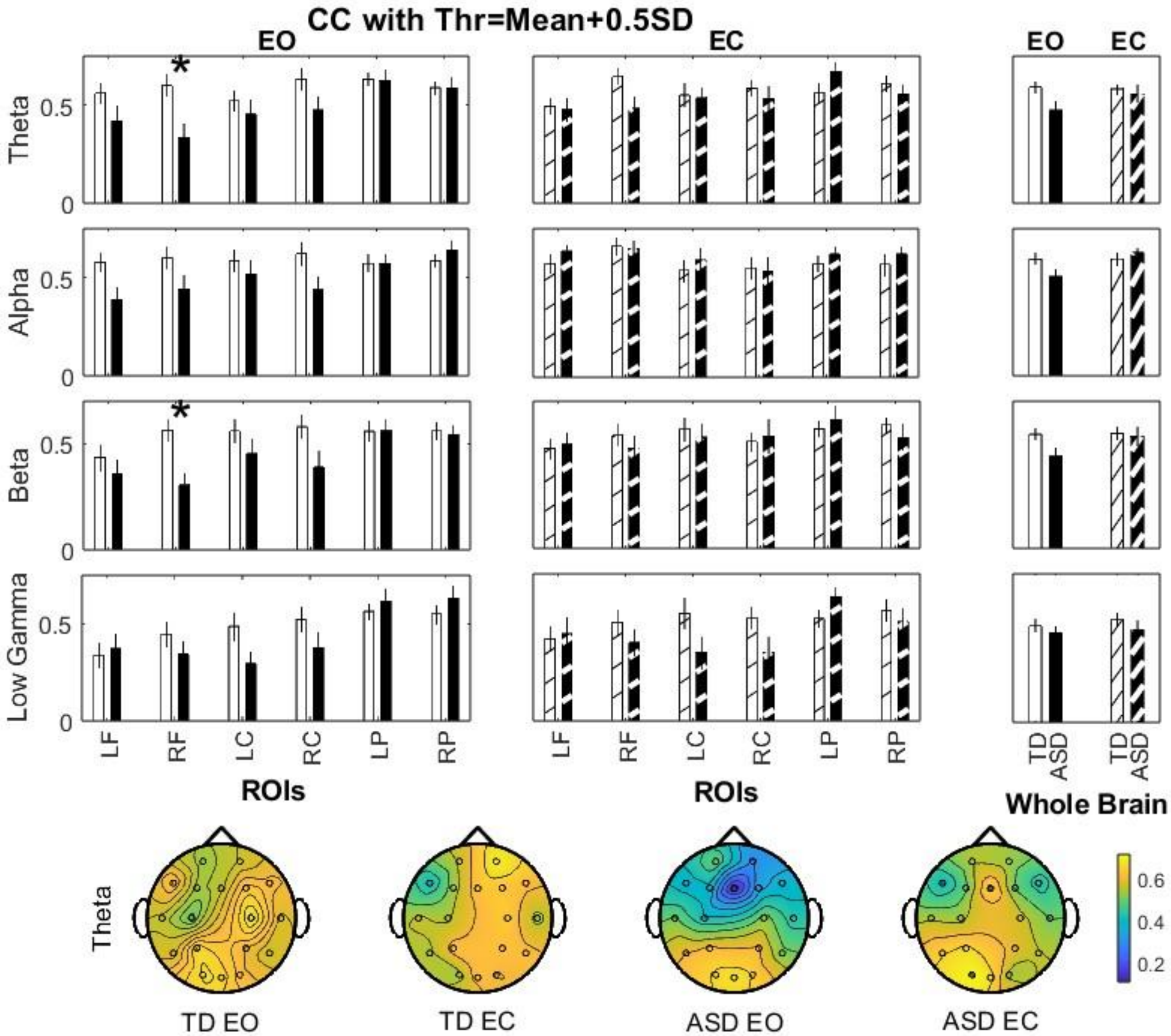

Fig 8. Clustering coefficient (CC) with Threshold=Mean+0.5SD. Group comparison with EO (left panels) and EC (right panels). TD EO (white bar), TD EC (white bar with black slanted lines), ASD EO (black bar), ASD EC (black bar with white slanted lines). Bar graphs with Mean ± SEM. *p<0.0083 in ROIs and *p<0.025 in Whole Brain. Topoplots for each group and each condition at theta band.

| | Condition | Group | ROI | Interaction |
|---|---|---|---|---|
| Theta Band | F(1,40)=0.59, p>0.05 | F(1,40)=1.93, p>0.05 | **F(4.1,163.4)=5.45, p=3.42e-4** | None (p>0.05) |
| Alpha Band | F(1,40)=3.21, p>0.05 | F(1,40)=0.57, p>0.05 | F(4.0,158.4)=1.08, p>0.05 | None (p>0.05) |
| Beta Band | F(1,40)=2.53, p>0.05 | F(1,40)=1.58, p>0.05 | **F(4.0,157.9)=3.54, p=8.74e-3** | None (p>0.05) |
| Low Gamma Band | F(1,40)=0.45, p>0.05 | F(1,40)=1.10, p>0.05 | **F(4.2,168.3)=5.85, p=1.49e-4** | **Cond*ROI [F(4.2,168.3)=2.82, p=2.44e-2]** |

Table 7. ANOVA for Clustering coefficient (CC) with Threshold=Mean+0.5SD in ROI.

*3.4.2. Network efficiency*

We also considered two different threshold values. While there were not enough statistical interactions between Group, Condition, and ROI for both threshold values, we conducted exploratory investigations into how opening and closing eyes affected neural activity differently within each group and condition in each region. For the threshold value equal to the mean of $\gamma$ from all control participants, efficiency was higher in TD than ASD in the right frontal region at beta and low gamma bands and in the left central region at the low gamma band in the EO condition (Fig. 9). There were no differences in the EC condition.

While there were not enough statistical interactions between Group and Condition for the whole brain consideration, we conducted exploratory investigations into how opening and closing eyes affected neural activity differently within each group. Global efficiency EF was significantly higher in TD than ASD at both beta band and low gamma band in EO condition (Fig. 9). There were no differences in EC condition. Closing the eyes elevated global efficiency EF in ASD (but not in TD) at the beta band (Supplemental Fig. 10).

For a higher threshold value equal to the mean + 0.5SD of γ from all control participants, there were no significant differences in the efficiencies in any of the conditions, groups, or regions (Supplemental Fig. 11 and 12).

While there were not enough statistical interactions between Group and Condition for the whole brain consideration, we conducted exploratory investigations into how opening and closing eyes affected neural activity differently within each group. Global efficiency EF was higher in TD than ASD for all regions at low gamma band in EO condition (Supplemental Fig. 11). Closing the eyes elevated global efficiency EF in ASD (but not in TD) at alpha band (Supplemental Fig. 12). Please refer to Table 8 (and Supplemental Table 10,11,12) for detailed statistical results.

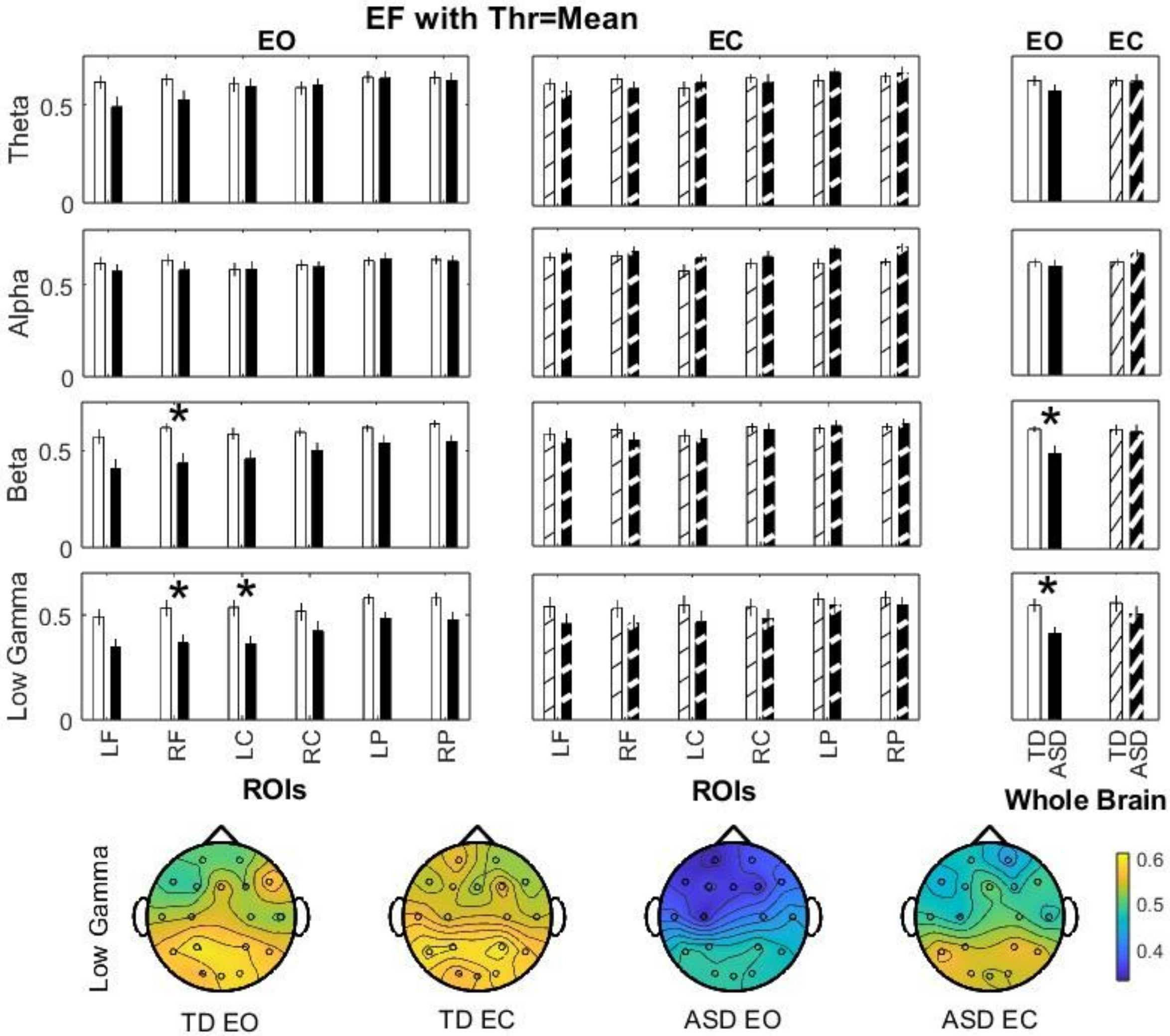

Fig 9. Efficiency with Threshold=Mean. Group comparison with EO (left panels) and EC (right panels). TD EO (white bar), TD EC (white bar with black slanted lines), ASD EO (black bar), ASD EC (black bar with white slanted lines). Bar graphs with Mean ± SEM. *p<0.0083 in ROIs and *p<0.025 in Whole Brain. Topoplots for each group and each condition at low gamma band.

| | Condition | Group | ROI | Interaction |
|---|---|---|---|---|
| Theta Band | F(1,40)=0.93, p>0.05 | F(1,40)=0.31, p>0.05 | **F(3.5,139.8)=6.76, p=1.19e-4** | None (p>0.05) |
| Alpha Band | F(1,40)=3.10, p>0.05 | F(1,40)=0.23, p>0.05 | **F(3.6,144.4)=3.62, p=9.82e-3** | None (p>0.05) |
| Beta Band | **F(1,40)=5.42, p=2.51e-2** | F(1,40)=3.55, p>0.05 | **F(3.4,134.9)=7.68, p=4.32e-5** | **Cond*Group [F(1,40)=5.48, p=2.44e-2]** |
| Low Gamma Band | **F(1,40)=3.88, p=5.59e-2** | **F(1,40)=4.40, p=4.24e-2** | **F(3.7,148.6)=7.27, p=3.75e-5** | None (p>0.05) |

Table 8. ANOVA for Efficiency with Threshold=Mean in ROI.

## 4. Discussion

The goal of the current study was to use dynamical system analysis to connect the measures of resting state EEG and behavior in young autistic adults in order to develop an integrative understanding of functional connectivity, and, in the global sense, neural control of behavior. We aimed not to validate specific measures, but rather to explore the neural dynamics using intentionally selected and newly developed measures that could characterize fine-grained temporal organization of brain networks that could be connected to behavioral measures. Below, we summarize our observations, discuss the functional network activity implications, and conclude with a discussion of potential hypotheses that may be suggested by our observations.

### 4.1 Summary of the Results

The behavioral measures indicated that ASD participants overall had higher social anxiety and social anhedonia in comparison to those of TD participants. The ASD participants also scored high on the AQ-10, confirming previous clinical autism diagnosis at the time of EEG recording. One typical concern with clinical research in autistic populations has to do with small sample sizes. In the present study, this concern is mitigated by high homogeneity among the participants who were high-functioning (low support needs) college students with independent diagnoses confirmed at the time of EEG recording using AQ-10.

In agreement with previous work, commonly used spectral measures do not indicate substantial differences between TD and ASD groups (Li et al., 2022). Statistical analysis also does not reveal any differences between ASD and TD in either of the brain regions or the brain as a whole in any of the considered spectral bands for EO or EC conditions. Of course, closing the eyes largely leads to broad alpha potentiation (especially posteriorly), but it happened in both groups. Considering the data from a different angle, specifically the differences induced by closing the eyes, there were several statistically significant differences between the two groups such as the elevation of theta and beta PSDs in different regions between ASD and TD, and the elevation of alpha oscillatory power in ASD (although these might not be robust phenomenon with clear functional implications). Additionally, while PSD and oscillatory power are related measures, the latter is essentially a power spectral density after the 1/f component is removed (whose properties are discussed later in the text).

This picture of the broad similarity of spectral characteristics between TD and ASD is in line with earlier studies (Li et al., 2022), emphasizing the question of neurophysiological differences underlying the distinct ASD social, cognitive and behavioral profile. The measures of network connectivity and dynamical system organization discussed below are more successful at capturing the differences in neurophysiology between two populations.

The two measures discussed next are best summarized as measures of complexity, as they are both related to the irregularity (or noise) in the structure of the signals that are informative regarding the complexity of the organization of the underlying system. The first measure is the 1/f exponent, which we referred to as one of the spectral measures in *Methods and Results*, and which characterizes 1/f non-oscillatory background. A larger exponent (that is, the larger power of *f* in *1/f* decaying spectral profile) corresponds to a less noisy and more regular signal. The second measure is the maximal Lyapunov exponent. Lyapunov exponent characterizes the rate of divergence of the nearby trajectories in the reconstructed phase space. In other words, it tells us how quickly two very similar states in the network will grow apart in time. A larger Lyapunov exponent indicates faster divergence and, thus, less predictable, more irregular system dynamics. These measures are not equivalent to each other, and are based on different time-series analysis approaches (spectral analysis vs. phase space reconstruction) but are characteristics of similar behaviors in neural dynamics. 1/f exponent did not show significant differences between the groups in either of the areas or conditions. However, closing the eyes showed a significant elevation of *1/f* exponent in the ASD group, pointing to the less noisy neural activity in this state in ASD (but not in TD) participants. The Lyapunov exponent indicated more robust changes being significantly higher in ASD, pointing to less stable (faster diverging) neural dynamics in ASD. This may reflect a form of neural instability or unpredictability that can underlie behavioral characteristics such as heightened sensory reactivity or variable responses to similar social contexts. Rather than being globally chaotic, this increased temporal divergence may indicate difficulty in maintaining stable network states during rest. When considering the effect of closing the eyes on the state of the brain at rest, Lyapunov exponent increased in both groups across all brain regions but showed more region-specific increases in the neurotypical group.

Our analysis also considered two synchrony-related measures. One measure is the phase-locking strength index; larger values of this index indicate stronger synchrony, i.e. stronger temporal coordination of oscillatory activity. The other measure is the desynchronization ratio. The larger value of the desynchronization ratio indicates the stronger prevalence of short desynchronization events in partially synchronized oscillatory dynamics. The desynchronization ratio may vary independently of the phase-locking index (Ahn et al., 2011, 2014; Ahn and Rubchinsky, 2017; Nguyen and Rubchinsky, 2021, 2024), as the same synchrony strength may be achieved via different numbers of shorter or longer desynchronizations. The phase synchrony index was higher in the EO condition in the TD group (no difference in the EC condition); the desynchronization ratio was higher in the TD group as well. For the whole brain, the effect is observable for higher frequency (low gamma); for the region-specific analysis, the effect is localized to the left central region at the low gamma band (possibly driving the whole brain effect in the low gamma range). While closing the eyes did not lead to significant changes in the TD group, in the ASD group this led to synchrony becoming stronger, and desynchronizations becoming shorter (as indicated by higher DR). While on the whole synchronization measures differences were not consistently significant across all comparisons, the noted combinations of higher DR and increased local phase synchrony in ASD group support the hypothesis that long-range neural communication is disrupted in ASD. Reduced global synchrony could impair coordinated activity across brain areas involved in multisensory integration and social cognition—areas in which autistic individuals often show distinct behavioral profiles.

Finally, we considered two functional network measures, clustering and efficiency. Let us emphasize that while these are network measures, their networks are functional, not anatomical. So, while the recorded EEG signal is generated by the underlying anatomical structures, the properties of the architecture of the functional networks speak about neurophysiology rather than neuroanatomy. The Clustering coefficient (CC) measures the tendency to form locally dense and characterizes the robustness of a network. On the other hand, efficiency measures the reciprocity of the shortest path length and characterizes the efficiency of information transfer in the network. While the details of results depend on the threshold selection for functional network reconstruction, in the EO condition the ASD group showed functional networks that were less connected in both cluster and efficiency than in the TD group; this was especially evident for higher frequencies (beta and gamma). There were no significant differences in the EC condition. The effect of closing the eyes for the ASD group was broader functional connectivity (interconnectivity in alpha and efficiency in beta), with no significant changes in the TD group.

In terms of measures of neural dynamics, the most interesting differences between groups were found in the resting state with eyes open. While the eyes closed condition is, generally, regarded as less noisy (without blinks), it is also always accompanied by a strong posterior alpha component, which contaminates all cross-frequency measures, obliterating more subtle or transient differences. Thus, eye-open state analysis revealed differences between groups that were not observable in eyes-closed data. However, we may consider the eyes closing as a factor that elicits changes in brain dynamics in the ASD and TD groups. The unavoidably noisy data means that the statistically significant effect of changes induced by closing the eyes is not just the difference between the significant results between the EO and EC conditions in all possible cases. We can consider the effect of closing the eyes as the response of the neural circuits of the brain to a substantial perturbation (which leads to changes in sensory inputs, potentiation of alpha, reorganization of the brain activity, etc.) and elicits different responses in the ASD and TD groups.

## 4.2. Neurodynamics of autism

We interpret the summary of the results in the context of its potential network activity implications. While there were no significant differences between groups in the EC condition, ASD group data shows lower synchrony, lower desynchronization ratios, higher Lyapunov exponents, and lower clustering coefficients and network efficiency in the EO condition. This suggests that if sensory defensive behaviors such as closing of the eyes are unavailable, the temporal network organization of the autistic brain yields dynamic behavior that is more random and has less spatially and stably connected dynamics in comparison to those of neurotypical peers. This results in distinct modes of neural dynamics organization in autistic participants. As the participants are young adults, these modes appear to reflect life-long adaptations, evident both in resting-state neural activity and social cognitive processing strategies, as evidenced by behavioral data. The increased social anxiety and social anhedonia in ASD, as compared to TD participants, noted in behavioral results, might arise as a long-term strategic adaptation, if the sensorily-defensive mode of neural dynamics observed in the eyes-opened condition is adopted for everyday functioning. Disrupted functional connectivity in resting state has been previously noted for both autism (Simmons et al., 2012) and social anxiety disorder (Ding et al., 2011).

One of the differences observed when considering EEG data as reflective of the underlying dynamical systems model is that closing of the eyes makes the dynamical system more structured, regular, and coordinated in autistic participants, while the effect of closing the eyes was much less robust in neurotypical participants and point towards more irregular dynamics. Several parameters converge on this interpretation: measures of regularity, measures of functional network connectedness, and measures of synchrony. Importantly, these changes are observed at rest. The participants are not engaged in any task (except, of course, mind-wandering). So, the brain state in the resting state should be able to rearrange itself into task-performing mode. Closing the eyes should likely decrease external sensory drive in the rest state. Thus, it appears that the eyes-closed condition modulates autistic brain dynamics toward a configuration more similar to that of neurotypical participants in terms of increased regularity and connectivity. These findings raise the possibility that atypical neurodynamic patterns in ASD, including increased temporal instability or reduced network flexibility, may contribute to altered sensory experiences or reduced adaptive ability. Importantly, however, individual neurobiological factors also affect one's interaction with the social environment. A growing body of work indicates that social withdrawal in autism may be a consequence of negative experiences. For example, Crompton et al. (2025) demonstrates that autistic individuals show intact information transfer and interpersonal rapport in interactions with other autistic participants, while in mixed neurotypical-autistic dyads the rapport is markedly lowered. These results bring focus to the relational nature of social difficulties, as driven by mismatches in neurotype and communication style. Individual heightened sensory reactivity may make autistic individuals more vulnerable to adverse social environments, and, over time, give rise to social withdrawal patterns.

Finally, we'd like to make a note of the spatial organization of the observed changes. The changes appear to be more consistent over the frontal cortex, or executive regions (rather than posterior cortical regions), which would be expected to control the among-network switching that is characteristic of resting state brain activity (Betzel et al., 2012; Kitzbichler et al., 2015). This might suggest that executive control of distal brain areas needed for complex social communication is more difficult for ASD participants: it is not as linearly related to the processing of the communicative signal as in neurotypical individuals.

Taken together, these exploratory EEG findings provide converging evidence for a model in which ASD is characterized by more locally stable but less dynamically flexible global brain activity patterns, consistent with observed behavioral traits such as insistence on sameness, sensory sensitivities, and context-inflexible social engagement. Anterior regions showed the most consistent findings for ASD group in our dataset, possibly reflecting differences in executive function and language-related circuits that are commonly implicated in ASD. Future research with larger and more stratified samples could validate these interpretations and link them directly to task performance or ecological behavioral outcomes.

### 4.3 Conclusions and implications

Applying a complex systems approach to understanding the neural bases of autism holds potential for addressing socially relevant questions about communication and behavior. Previous work suggests autistic individuals may be particularly sensitive to input signals due to distinct temporal patterns in brain networks (Malaia et al., 2020); computational studies provide further evidence for this possibility (Nguyen & Rubchinsky, 2024).These findings could reflect an adaptive inhibition strategy in autism. Even though the present study is very remote from any therapies or interventions in ASD, it suggests that elucidating such neural mechanisms underlying the autism phenotype may optimize the support and accommodations provided to autistic individuals across the lifespan. Moreover, it could enhance societal understanding of the bidirectional nature of communication differences between neurotypical and neurodivergent populations (Cockerham & Malaia, 2016; Davis & Crompton, 2021). A key limitation of the current study is the specificity of participant profile: all ASD individuals were cognitively able, high-functioning students enrolled in postsecondary education. While this allows for reduced heterogeneity in cognitive performance, it also limits the generalizability of the findings. Future research should examine whether the neural dynamics patterns observed here extend to other groups on the autism spectrum, including individuals with higher support needs or co-occurring intellectual or developmental conditions.

**Data Availability**

Data will be made available upon reasonable request from qualified researchers with permission from the IRB.

**Acknowledgements**

Supported by AMS-Simons Grant for PUI (SA). We are also thankful to the participants who took part in this study.

**Declaration of Competing Interest**

The authors declare that they have no known competing financial interests or personal relationships that could have appeared to influence the work reported in this paper.

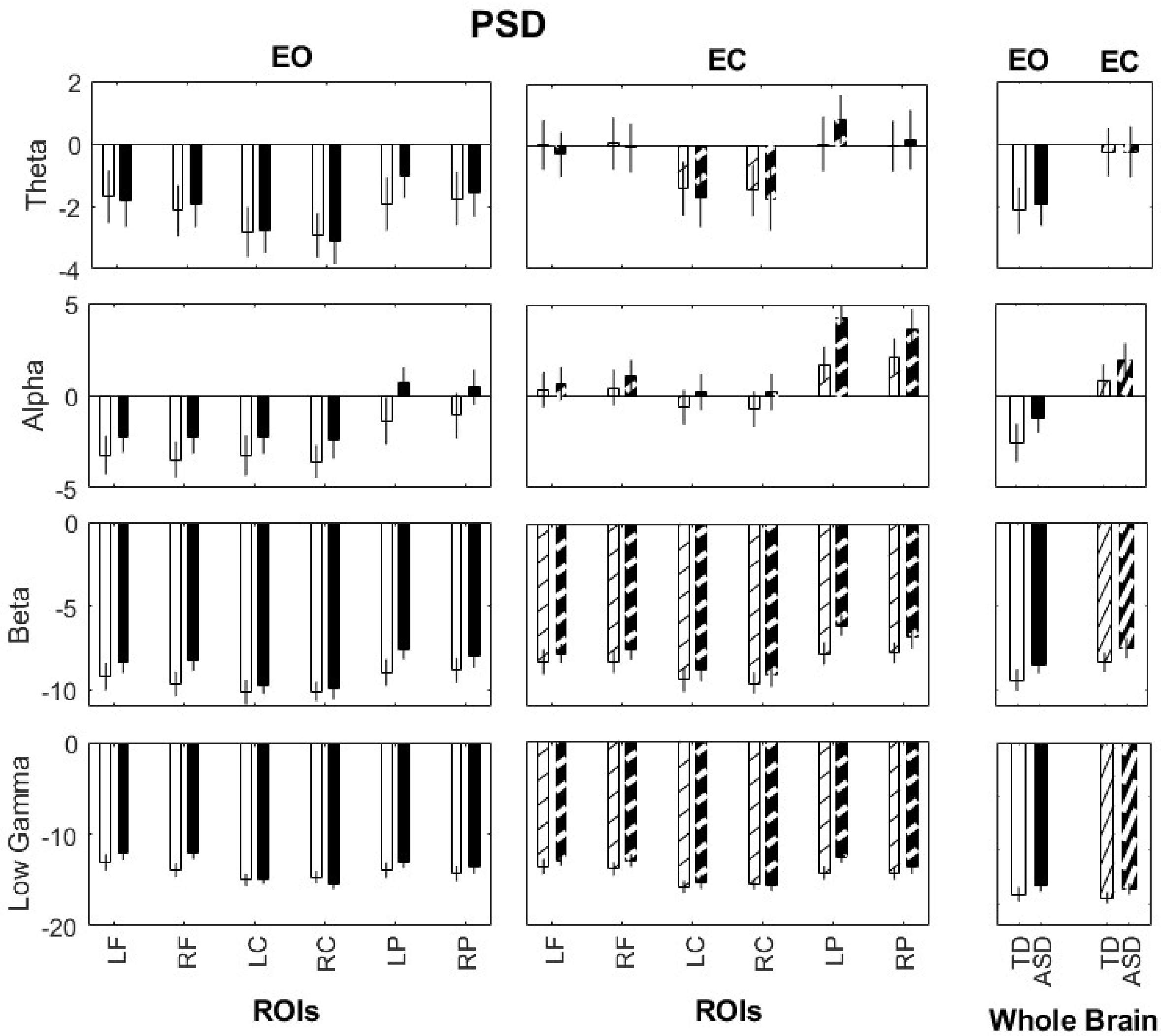

Supplementary Figure 1.

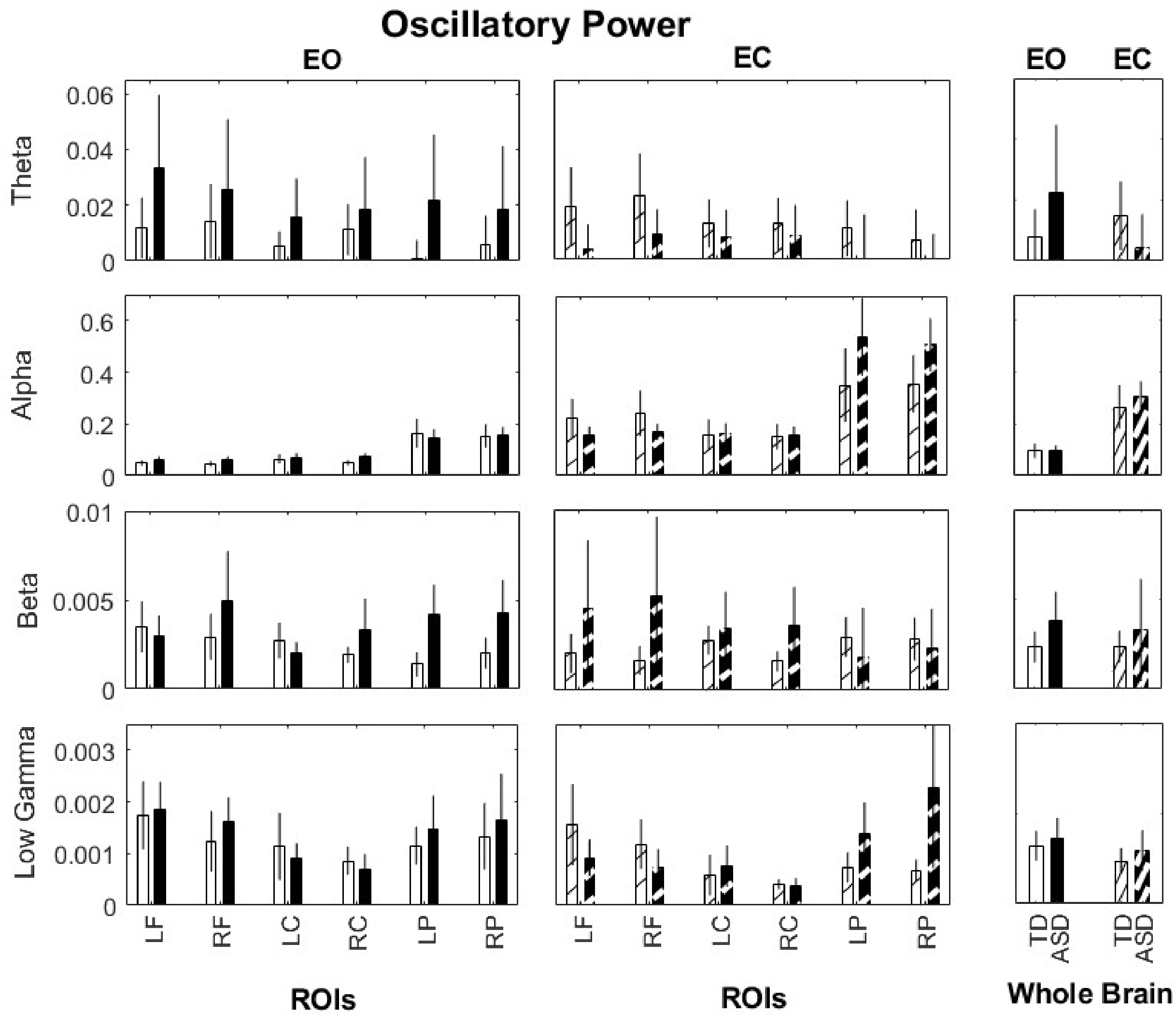

Supplementary Figure 2.

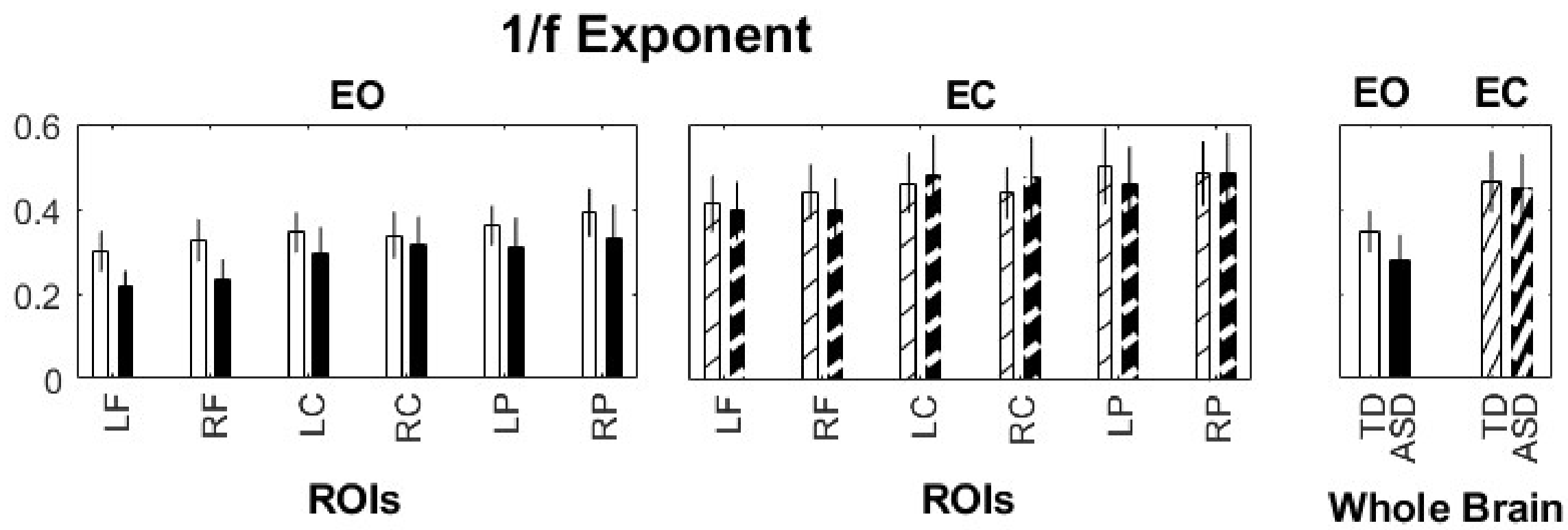

Supplementary Figure 3.

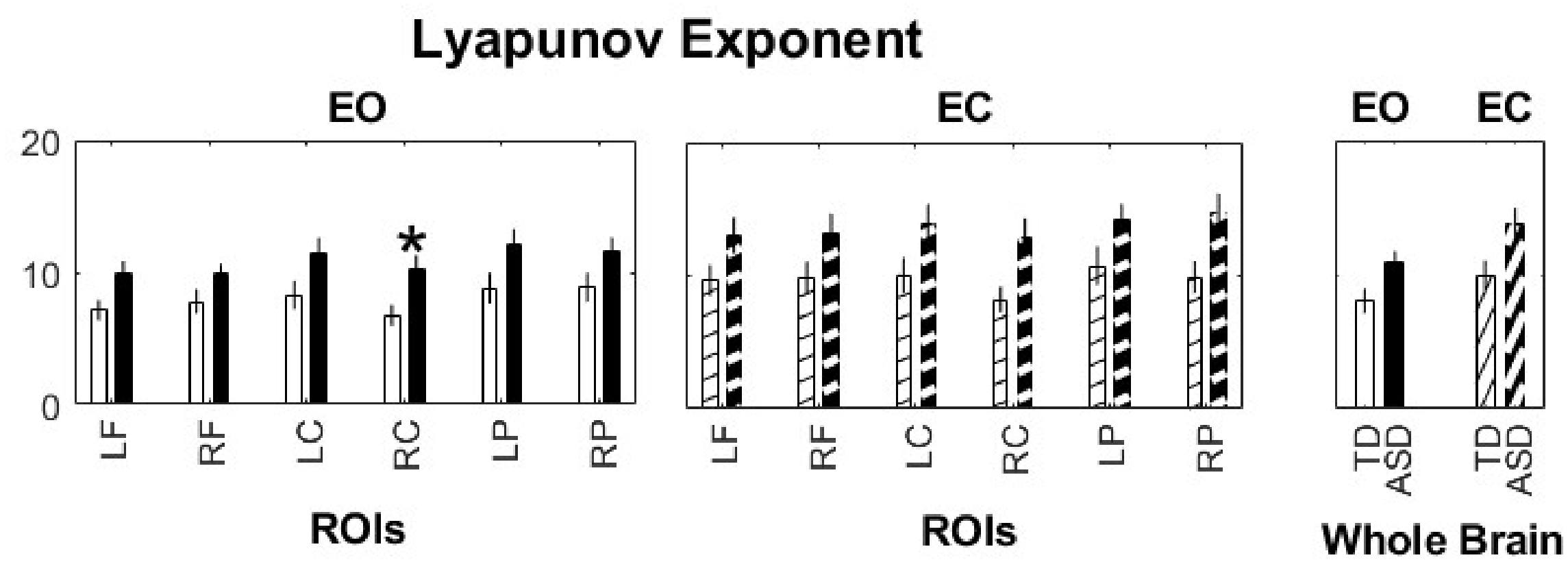

Supplementary Figure 4.

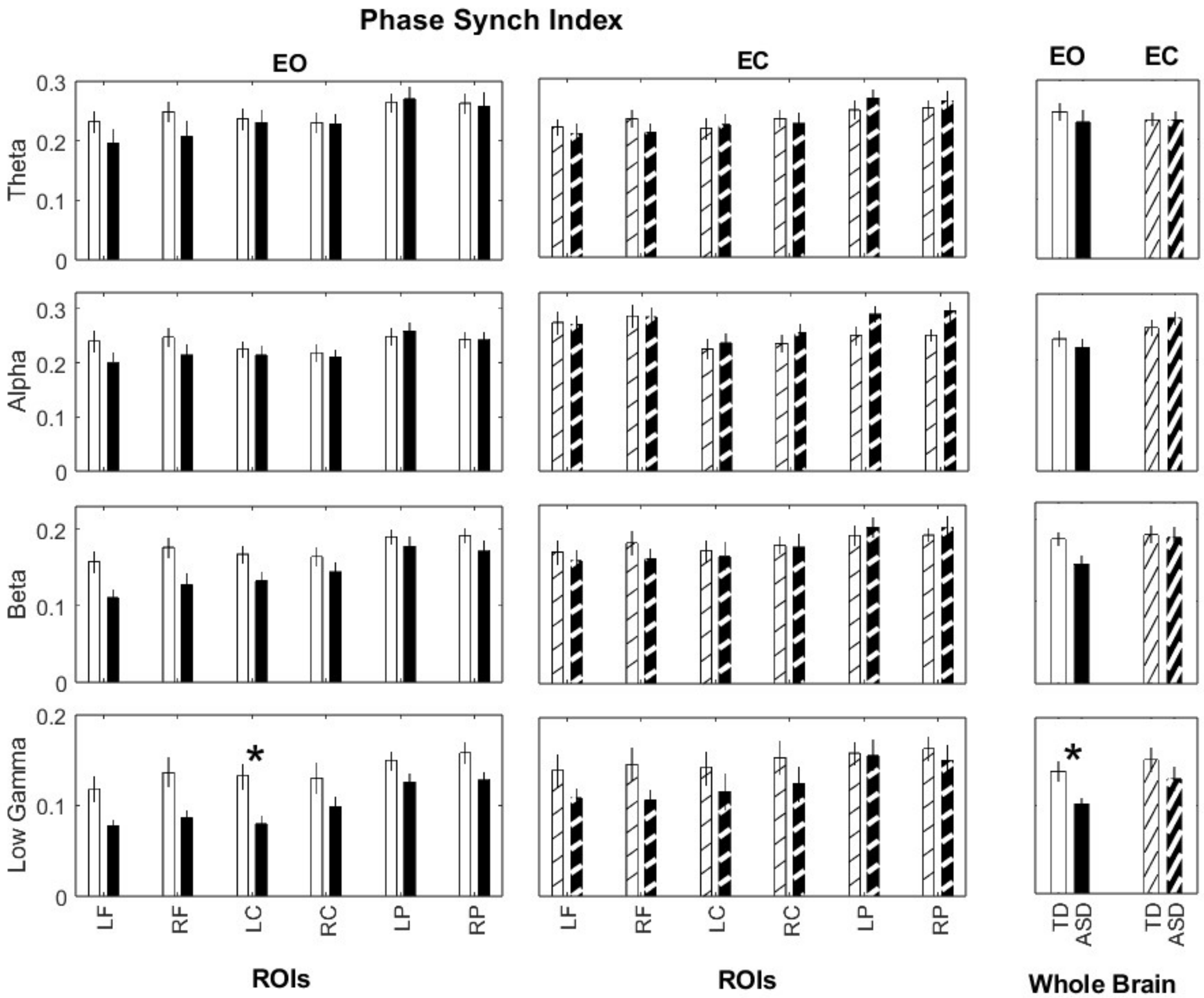

Supplementary Figure 5.

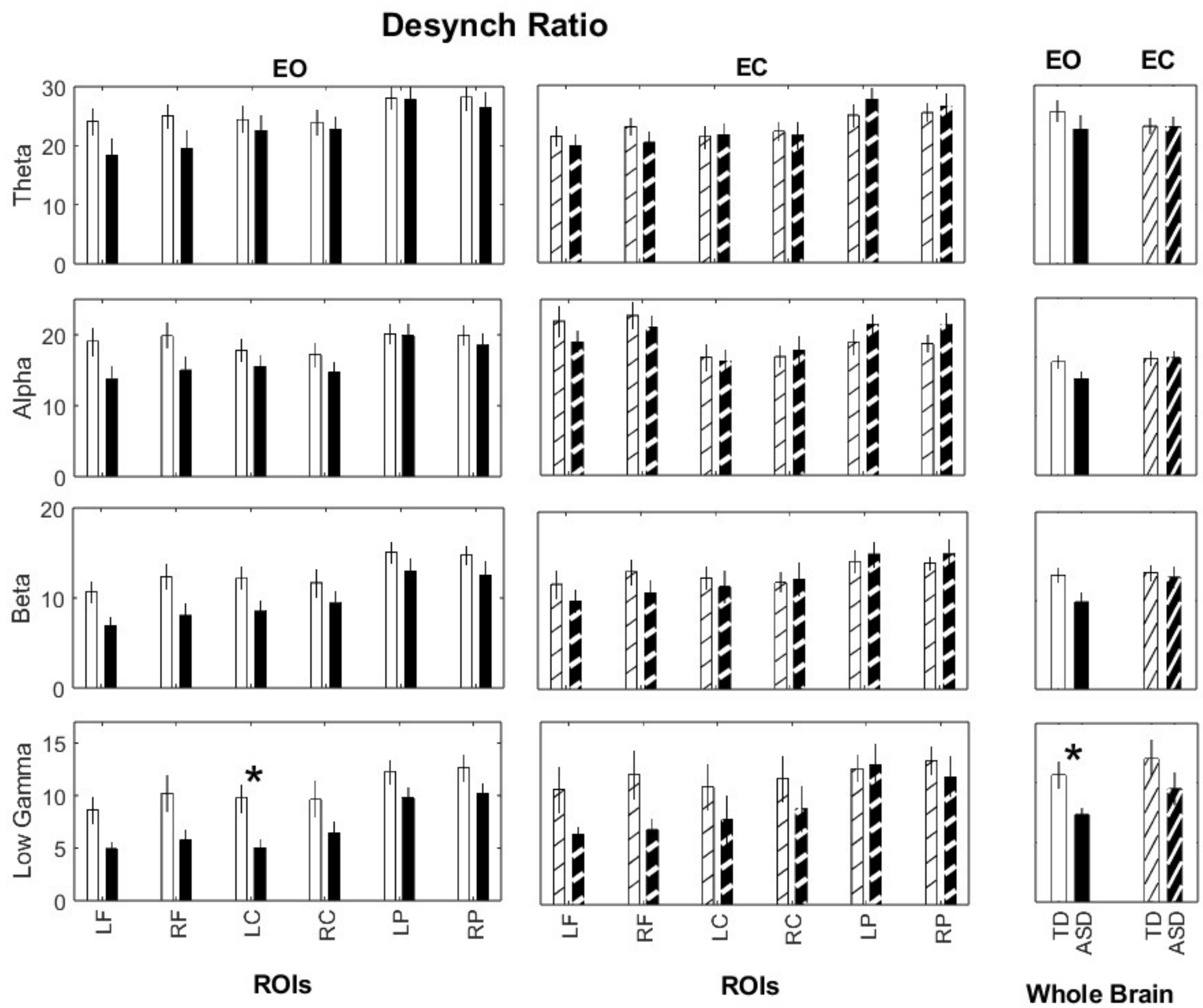

Supplementary Figure 6.

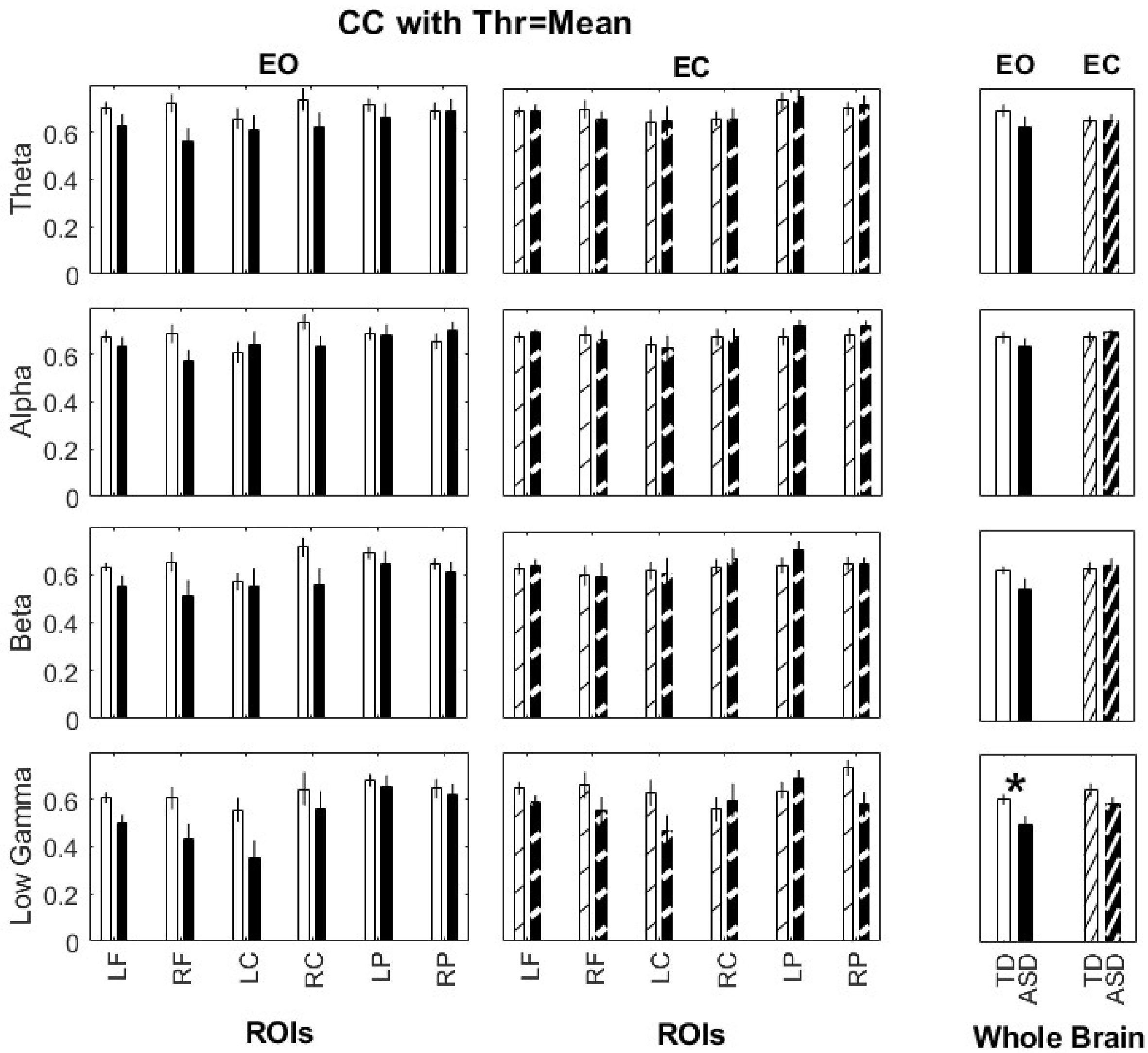

Supplementary Figure 7.

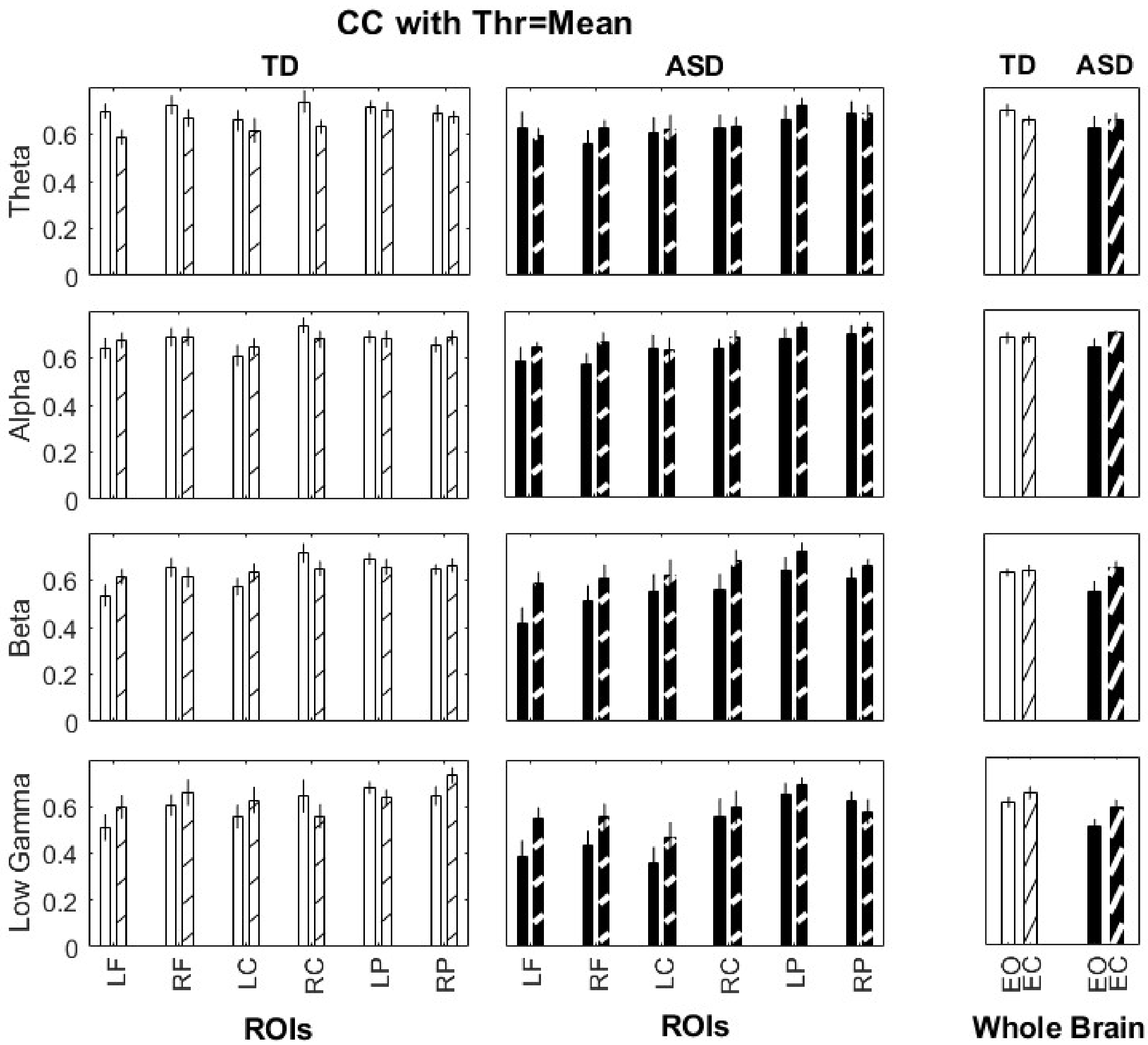

Supplementary Figure 8.

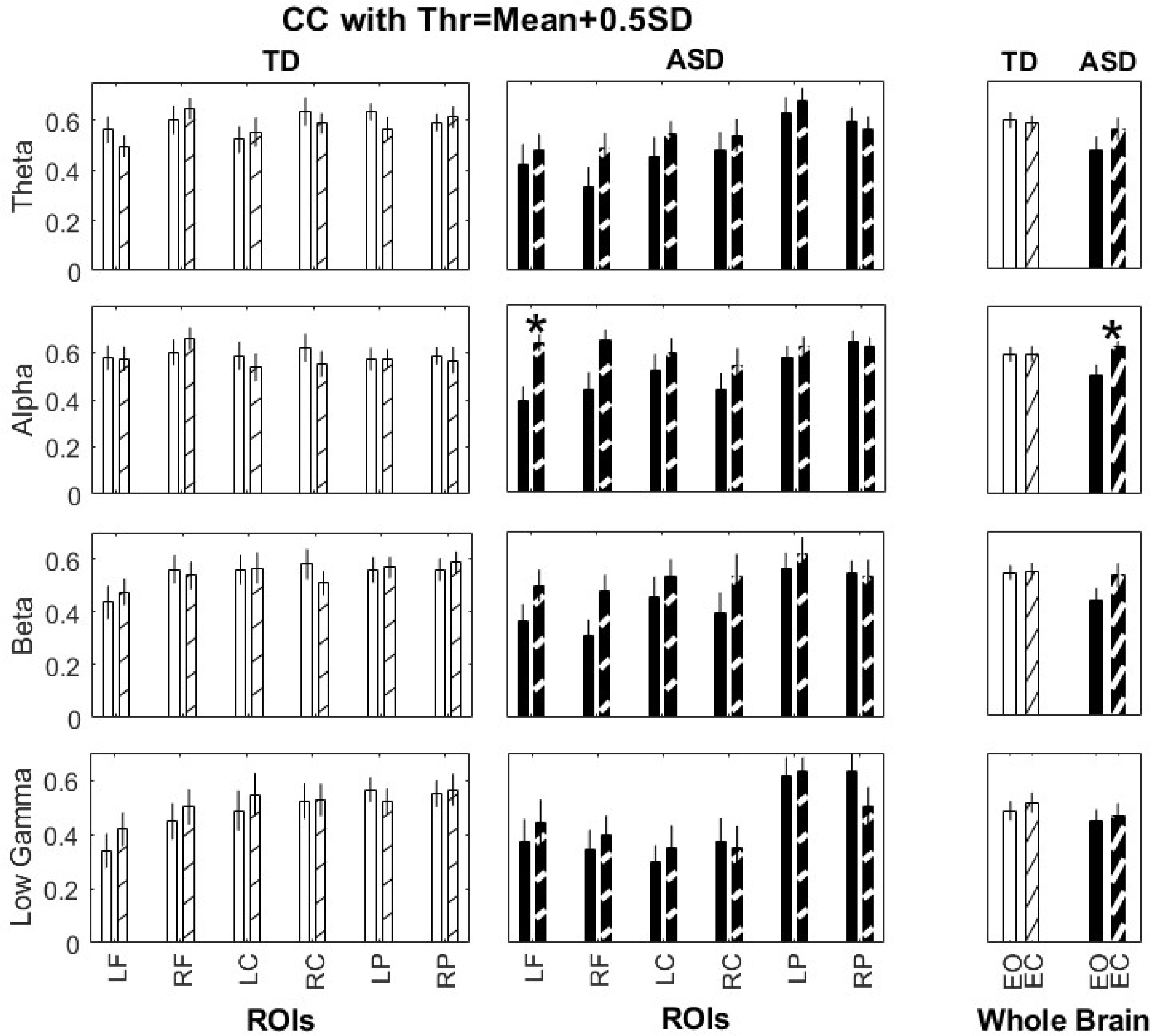

Supplementary Figure 9.

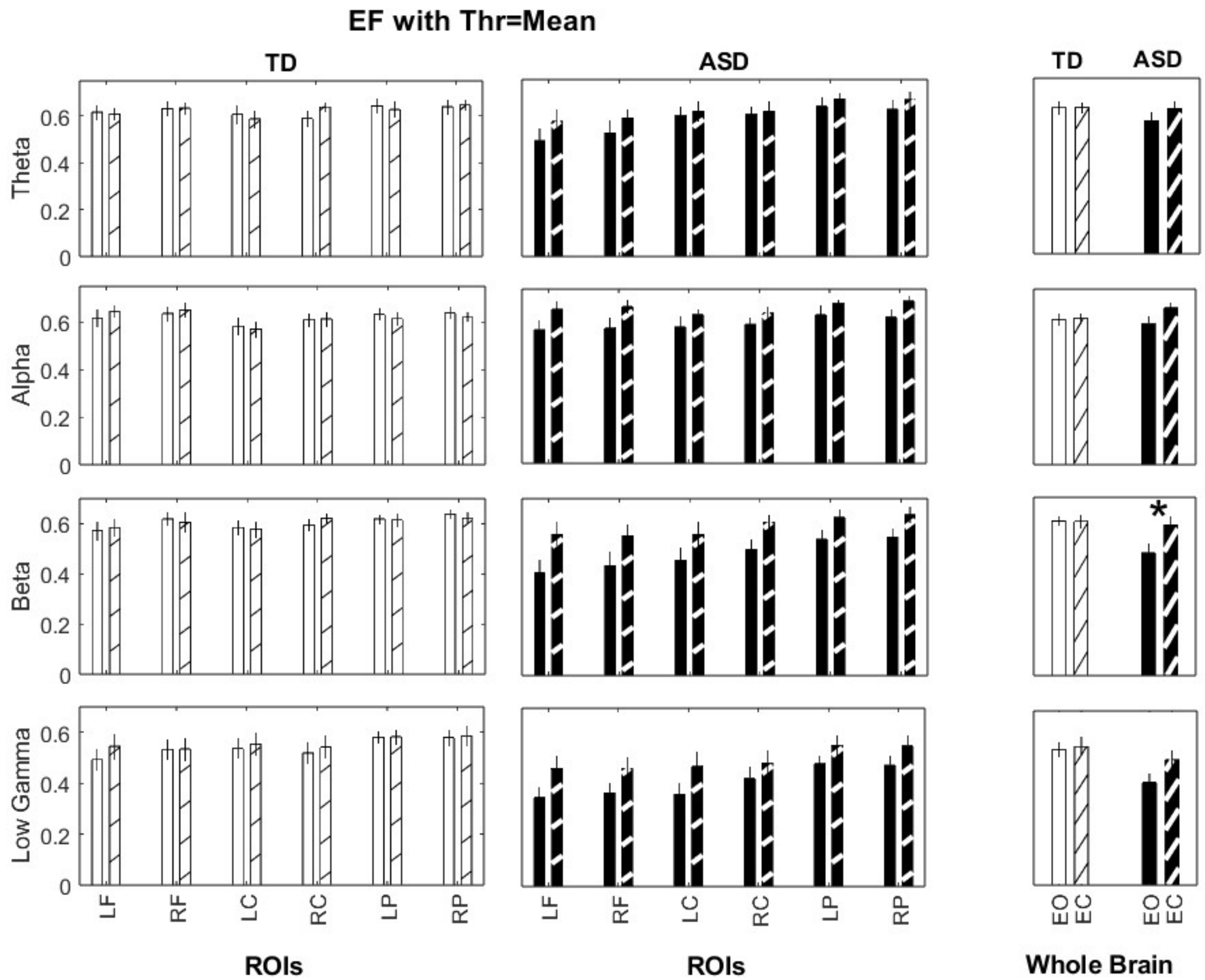

Supplementary Figure 10.

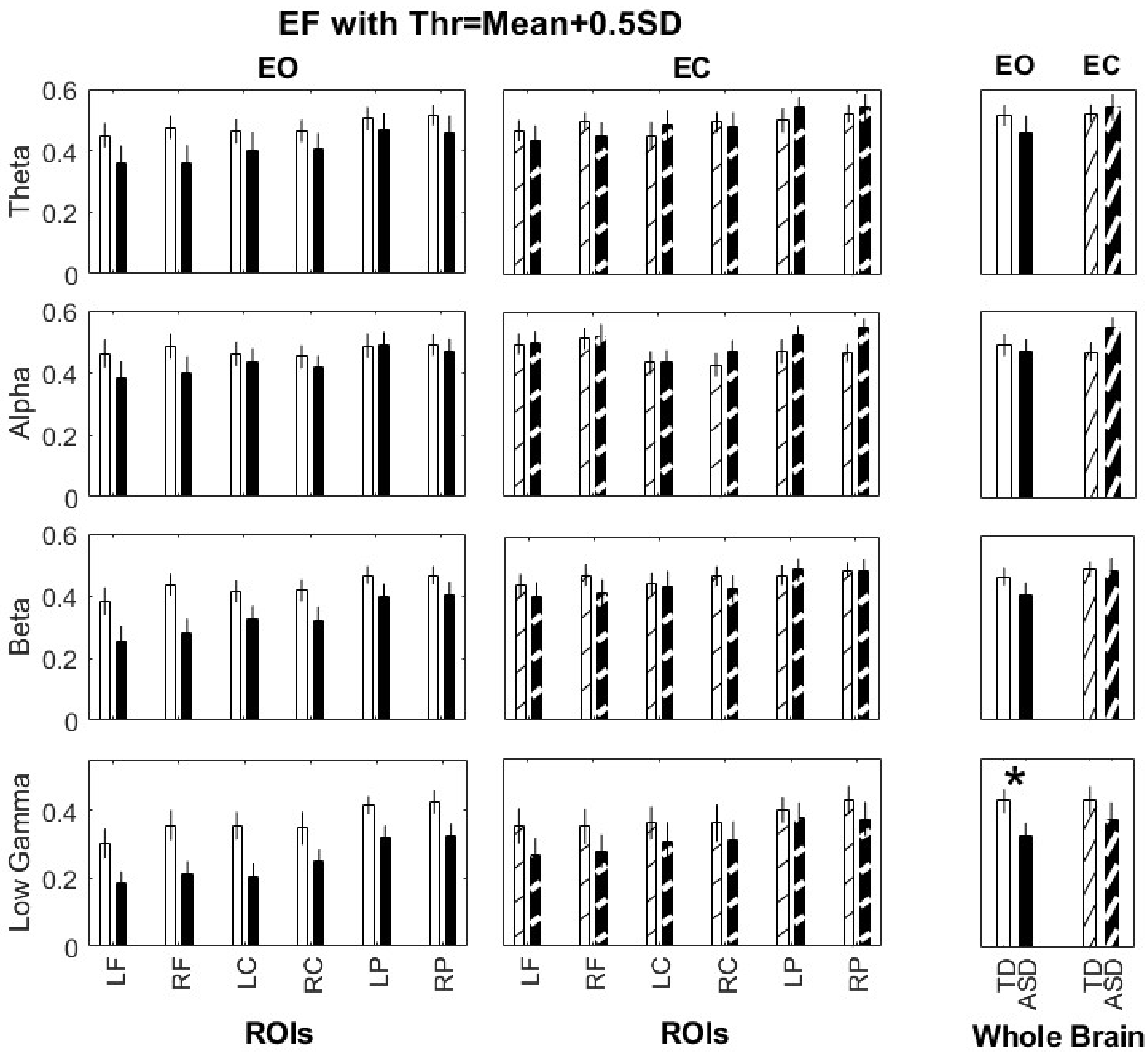

Supplementary Figure 11.

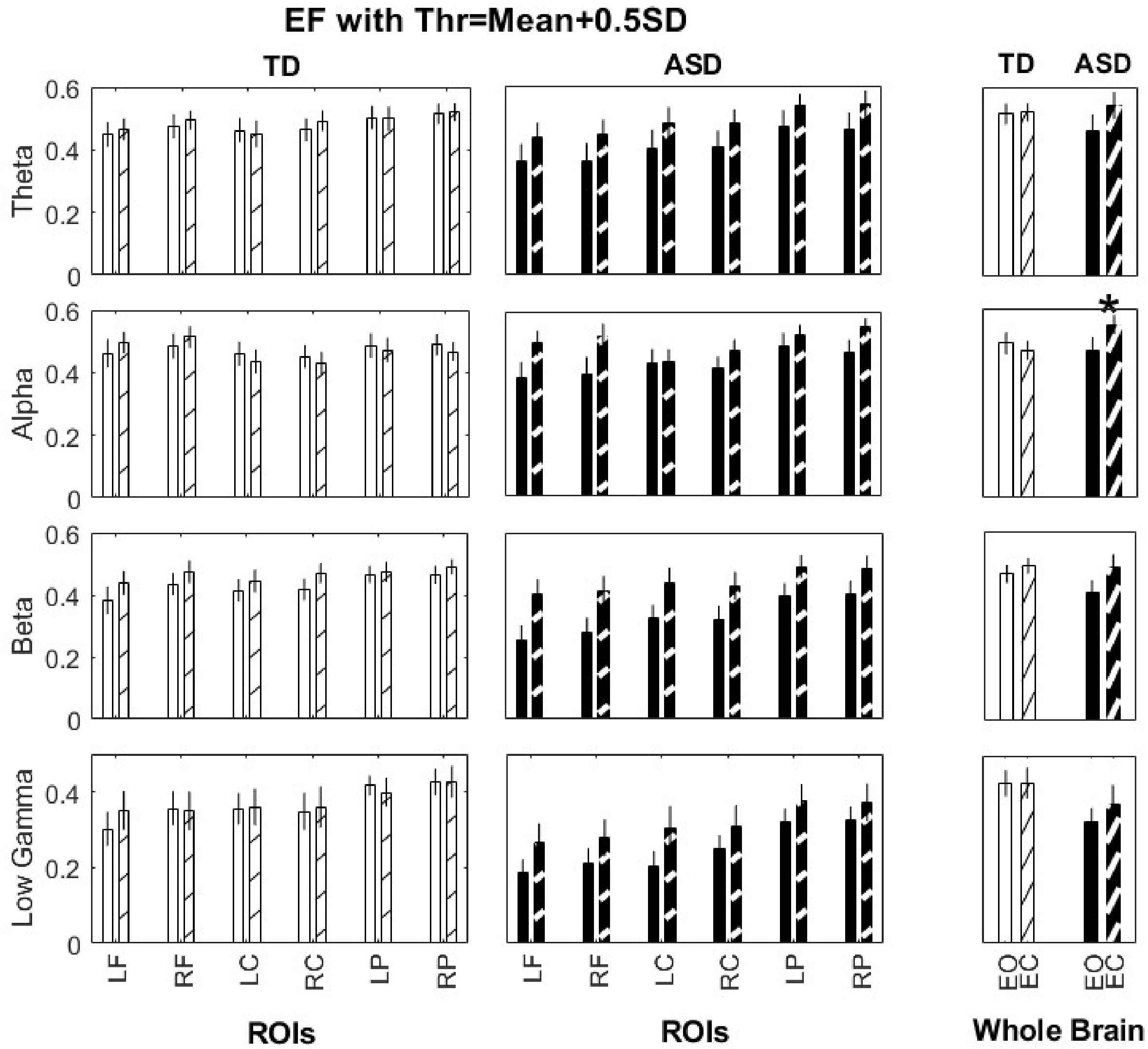

Supplementary Figure 12.